\documentclass[aps,prd,superscriptaddress,nofootinbib,11pt]{revtex4}
\usepackage[english]{babel}
\usepackage[utf8]{inputenc}
\usepackage{graphicx}   
\usepackage{slashed}
\usepackage{epstopdf}
\usepackage{verbatim}   
\usepackage{color}      
\usepackage{subfigure}  
\usepackage{multirow}
\usepackage{hyperref}   
\usepackage{float}
\usepackage{epsfig,rotating}
\usepackage{amsmath,amssymb}
\usepackage{dsfont}
\restylefloat{table}
\numberwithin{equation}{section}

\newcommand{\vp}{\vec{p}}
\newcommand{\vq}{\vec{q}}
\newcommand{\vk}{\vec{k}}

\newcommand{\be}{\begin{equation}}
\newcommand{\ee}{\end{equation}}
\newcommand{\bea}{\begin{eqnarray}}
\newcommand{\eea}{\end{eqnarray}}

\newcommand{\ket}[1]{|#1\rangle}
\newcommand{\bra}[1]{\langle#1|}

\newcommand{\ma}{\mathcal A}
\newcommand{\wpo}{\widetilde{p}_0}

\begin{document}
\title{Dynamics of relaxation and dressing of a quenched Bose polaron.}
\author{Daniel Boyanovsky}
\email{boyan@pitt.edu} \affiliation{Department of Physics and
Astronomy, University of Pittsburgh, Pittsburgh, PA 15260}
\author{David Jasnow}
\email{jasnow@pitt.edu} \affiliation{Department of Physics and
Astronomy, University of Pittsburgh, Pittsburgh, PA 15260}
\author{Xiao-Lun Wu}
\email{xlwu@pitt.edu} \affiliation{Department of Physics and
Astronomy, University of Pittsburgh, Pittsburgh, PA 15260}
\author{Rob C. Coalson}
\email{coalson@pitt.edu} \affiliation{Department of Chemistry, University of Pittsburgh, Pittsburgh, PA 15260}
\date{\today}

\begin{abstract}
We study the non-equilibrium dynamics of relaxation and dressing of a mobile impurity suddenly immersed--or quenched-- into a zero temperature homogeneous Bose Einstein condensate (BEC) with velocity $v$. A many body generalization of Weisskopf-Wigner theory is implemented to obtain the impurity fidelity,    reduced density matrix and entanglement entropy.
The dynamics depend crucially on the Mach number $\beta =v/c$,  with $c$ the speed of sound of superfluid phonons and features many different time scales. Quantum Zeno behavior at early time is followed by non-equilibrium dynamics determined by Cerenkov emission   of long-wavelength phonons for $\beta >1$ with a relaxation rate $\Gamma_p \propto (\beta-1)^3$.  The polaron dressing dynamics    \emph{slows-down} as $\beta \rightarrow 1$  and is characterized by  power laws $t^{-\alpha}$ with different exponents for $\beta \lessgtr 1$. The asymptotic entanglement entropy features a sharp discontinuity and   the residue features a cusp  at   $\beta =1$. These  non-equilibrium features suggest   \emph{universal} dynamical critical phenomena near $\beta \simeq 1$, and are  a direct consequence of the linear dispersion relation of long wavelength superfluid phonons. We conjecture on the emergence of an asymptotic dynamical attractor with $\beta \leq 1$.
\end{abstract}

\keywords{}

\maketitle

\section{Introduction.}

The original concept of the polaron by Landau\cite{landau} and Pekar\cite{pekar} describing the ``dressing'' of a charged particle by the polarization cloud in a polar material and moving together as a \emph{quasiparticle} in the medium\cite{leelowpines}, has become a paradigm of quantum many body physics\cite{frohlich1,feynman,devreese,mahan,bruun}.

Ultracold atoms provide an arena to study experimentally the main concepts of impurity dressing by the environmental degrees of freedom with unprecedented control over the various parameters and couplings\cite{bloch}. The emergence of quasiparticles from the dressing of an atomic impurity in a Fermi sea, namely a Fermi polaron, was experimentally observed by immersing a single atom in   imbalanced Fermi gases\cite{zwier1,zwier2,part,sommer,zaca}. Impurities immersed in a Bose-Einstein condensate (BEC)   become Bose polarons\cite{kalas,tim,tempe}. The observation of Bose polarons has been reported by radio frequency spectroscopy of ultracold bosonic $^{39}K$ atoms\cite{nils}, and for $^{40}K$ impurities in an ultracold atomic gas of $^{87}Rb$ \cite{jin}. The quasiparticle properties of the Bose polaron, such as the effective mass and wave function renormalization (residue) are studied in ref.\cite{quasi} in a systematic perturbation theory in the impurity-(BEC) scattering length.     Bose polarons of a (fermionic) $^{40}K$ impurity immersed in a $^{29}Na$ (BEC) near quantum criticality have been studied in ref.\cite{yun} probing spectral properties of the dressed quasiparticle via locally resolved radiofrequency spectroscopy.
The experimental possibilities of studying Bose polarons as paradigmatic of  the dynamics of impurities in ultracold quantum gases has fueled recent theoretical investigations of its quasiparticle properties \cite{demler1,rath,sarma,casteels,palma,levinsen,ardila,giorgini,demler,yoshida,camacho,pohl}. In ref.\cite{demlerinter} it is proposed that many body interferometry may offer a direct pathway to access the dynamics of the polaron cloud.

More recently a novel instance of a polaron has been shown to emerge in the case of molecules with rotational degrees of freedom immersed in a superfluid, giving rise to an ``angulon'' quasiparticle\cite{lemeshko} with properties very similar to polaron states resulting from the many body dressing of  electrons in polar materials.

Understanding the \emph{non-equilibrium} dynamics of a mobile impurity in a background of a cold degenerate gas remains a challenging problem. In ref.\cite{ideal} the real time dynamics of an impurity in a trapped \emph{ideal Bose gas} was studied within a T-matrix expansion in the impurity-Bose gas interaction. The quantum dynamics of Bose polarons has been studied with a time dependent variational ansatz in refs.\cite{ansatz,sarma},   near a Feschbach resonance  in ref.\cite{demlerdyn} and implementing dynamical renormalization group concepts in ref.\cite{drgdemler}. The quantum kinetics of thermalization and cooling of Bose polarons was studied within the framework of a Boltzmann Eqn. in ref.\cite{kinetics}. More recently the non-equilibrium   dynamics of Bose polaron formation and decoherence has been studied via a quantum master equation\cite{nielsen} in the Born approximation to second order in the impurity-(BEC) coupling. One of the main results in this reference is that the polaron formation time depends strongly on the velocity of the impurity,  exhibiting a critical slowdown for velocity close to the speed of sound of excitations in the BEC\cite{nielsen}.

\vspace{1mm}

\textbf{Motivation and main results:} The non-equilibrium dynamics of an impurity suddenly immersed in a BEC and the formation and evolution of a Bose polaron continues to be an important theme in quantum many body physics with timely experimental realizations with ultracold quantum gases. Our study is motivated by the fundamental and overarching question of non-equilibrium dynamics of quasiparticle formation, the possibility of experimentally probing this dynamics\cite{cetina1,cetina2,nils,ansatz} and the wealth of dynamical phenomena revealed by previous studies in refs.\cite{ansatz,demlerdyn,drgdemler,nielsen}. We implement a many body generalization of the Weisskopf-Wigner method\cite{ww} ubiquitous in quantum optics\cite{qoptics} to study the time evolution of an initial state that describes an impurity suddenly immersed in a BEC. This is an example of a ``quench'' because the interaction between the impurity and the BEC is turned-on suddenly. These methods provide a  non-perturbative resummation in real time, are conceptually and technically different from previous approaches, and have not hitherto been applied to the polaron problem.  They provide complementary information on the dynamical time scales, from the early transient to the long time dynamics of relaxation and dressing.  A bonus of this method is that it allows us  to obtain the Loschmidt echo or ``fidelity'' of the impurity, to study the build-up of correlations between the impurity and superfluid phonons and yields the full quantum state in real time revealing the dynamics of formation of the polaron cloud. Tracing over the phonon excitations yields the  impurity \emph{reduced density matrix}    from which we can obtain the von Neumann \emph{entanglement entropy} as a measure of correlations between the impurity and the BEC.

The non-equilibrium dynamics depends crucially on the Mach number of the impurity $\beta =v/c$ with $v$ the impurity velocity and $c$ the speed of sound of superfluid phonons. We focus on the case $\beta \simeq 1$ as this case reveals \emph{universal} long time dynamics. Early transient dynamics feature   quantum Zeno behavior with a cross-over to a stretched exponential. The long time dynamics is very different depending on whether $\beta \lessgtr 1$. For $\beta >1$ there are two different processes: a:) relaxation  via Cerenkov emission of phonons with a rate $\Gamma_p \propto (\beta-1)^3$, b:) dressing by virtual phonons with asymptotic dynamics that features a power law $\propto t^{-3/2}$ on a time scale $\propto (\beta-1)^{-2/3}$ modulated by  oscillations with frequency $\propto (\beta-1)^2$ as a consequence of a threshold in the spectral density.  For $\beta \leq 1$ only dressing by virtual phonons is available with an asymptotic long time dynamics featuring a  power law $t^{-1/2}$ for $\beta =1$ and $t^{-2}$ on a time scale $\propto (\beta-1)^{-3/2}$ for $\beta <1$  confirming a slow-down of dressing dynamics as $\beta \rightarrow 1$.

 We show that unitarity relates fidelity decay to the emergence of impurity-BEC \emph{correlations} and the growth of entanglement entropy, whose asymptotic long time behavior features a sharp transition across $\beta =1$.

Several of these results are in agreement with those found in ref.\cite{nielsen} for the dynamics of a \emph{coherence}. We explain the agreement by showing   a direct relationship between the coherence defined in ref.\cite{nielsen} and the amplitudes of the time evolved quantum state. Taken together these results hint at a \emph{dynamical} critical behavior with $\beta$ playing a similar role to $T_c/T$ in a theory of critical phenomena. This critical behavior is \emph{universal} for   $\beta \simeq 1$ in the sense that the power laws and exponents are independent of couplings and masses and are solely a consequence of the linear dispersion relation of long-wavelength superfluid phonons.

We \emph{conjecture} that if the impurity is quenched into the BEC with $\beta \gg 1$, relaxational dynamics leads the polaron to a \emph{dynamical} attractor manifold where the effective Mach number is $\beta \leq 1$ and the quantum state is described by the impurity entangled with multi-phonon states.

The article is organized as follows: section (\ref{sec:model}) introduces the model. Section  (\ref{sec:ww}) develops the many body generalization of Weisskopf-Wigner theory, and section (\ref{sec:polaron}) applies this framework to study the time evolution of a quenched impurity.  Section (\ref{sec:dressing}) discusses in detail the dynamics of dressing.   Section (\ref{sec:info}) discusses unitarity and the entanglement entropy.
Section (\ref{sec:critical}) discusses  the long-time dynamics within the context of dynamical critical phenomena,  section (\ref{sec:coherence}) establishes a direct relation with the framework   of ref.\cite{nielsen} discussing similarities and differences in the results,  and  section (\ref{sec:discussion}) discusses the regime of validity of the main approximation, various related aspects of our study and the \emph{conjecture} of an asymptotic dynamical attractor. Section (\ref{sec:conclusions}) summarizes our conclusions. Several appendices are devoted to technical aspects.

\section{The model.}\label{sec:model}

We consider the dynamics of an impurity of mass $M$ immersed in a three dimensional homogeneous Bose condensed gas (BEC) at zero temperature, described by the total Hamiltonian
\be H= H_B+ H_i + H_I  \equiv H_0+ H_I \,, \label{totalH}\ee with
\bea  H_B  & = &  \sum_{\vk} E_k \, b^\dagger_{\vk}\,b_{\vk} ~~;~~ E_k = ck\sqrt{1+\frac{k^2}{k^{*2}}} \,, \label{HB} \\
H_i & = & \sum_{\vk}\epsilon_k \, C^\dagger_{\vk} C_{\vk} ~~;~~ \epsilon_k = \frac{k^2}{2M}\,, \label{Hi}\\
H_I & = & \sum_{\vk} \frac{V_k}{\sqrt{\Omega}}\,C^\dagger_{\vp-\vk}\,C_{\vp}\,\big(b^\dagger_{\vk}+b_{-\vk}\big) \,, \label{HI}\eea
where $\Omega$ is the quantization volume. $H_B$ describes the Bogoliubov excitations of the Bose condensed gas  with phonon speed of sound $c$ and $k^*  = 2 m c$ with $m$ the mass of the   particles in the Bose gas. $H_i$ is the impurity Hamiltonian and $H_I$, the interaction Hamiltonian,  describes a Fr$\ddot{o}$hlich model\cite{frohlich1}. The matrix element of the interaction is given by
\be V_k = U_0   \,\sqrt{\frac{n_0\,k^2}{2mE_k}}\,, \label{Vk}\ee where $U_0$ is a local interaction vertex and $n_0$ is the BEC condensate density. This interaction may be understood simply from an impurity-BEC density-density local interaction, namely
\be H_I = U_0 \int d^3x \Psi^\dagger(\vec{x})\Psi(\vec{x})\psi^\dagger(\vec{x})\psi(\vec{x}) \,, \label{Hpsis}\ee where $\Psi,\psi$ are the second quantized fields associated with the impurity and the BEC respectively. In the Bogoliubov approximation with $\psi(\vec{x}) = \sqrt{n_0}+\delta\psi(\vec{x})$, diagonalizing the quadratic form of the BEC Hamiltonian in terms of Bogoliubov coefficients and keeping only the cubic interaction term between the impurity and the Bogoliubov excitations, one arrives at $H_I$. We have neglected a mean field constant correction $\delta \epsilon \propto U_0 n_0$ to the impurity energy in  (\ref{Hi}),    because it is not relevant for the discussion. The validity of a Fr$\ddot{o}$hlich interaction Hamiltonian relies on a macroscopically large number of atoms in the BEC ground state and on weak coupling. Such interaction has been implemented in various references\cite{tempe,nielsen,ansatz}. A more consistent treatment in terms of the T-matrix   is provided in ref.\cite{nielsen} and  yields a similar result. The main physical phenomena discussed in this article are described by long-wavelength, low energy phonon excitations, justifying  \emph{a posteriori} the local approximation for the interaction between the impurity and the BEC.

We consider that the impurity is immersed in the zero temperature ground state of the BEC at time $t=0$ and follow the time evolution of this initial state. This corresponds to a \emph{quench} because it is equivalent to switching the interaction on at $t=0$.

We now introduce the effective coupling
\be \lambda^2 = \frac{U^2_0\,n_0}{(2\pi)^2\,2mc} \,,\label{lambda}\ee whose dimensions are $1/(mass)^2$. As   will be discussed in detail below, the effective \emph{dimensionless} coupling relevant for the low energy, long time dynamics is the dimensionless combination $ \lambda^2 M^2 $.

We   restrict our study strictly to \emph{weak coupling}, namely $\lambda^2 M^2 \ll 1$,  postponing to future work the extension to strong coupling.

\section{\label{sec:WW} Many Body generalization of Wigner-Weisskopf theory. }\label{sec:ww}

Our main goal in this article is to study the time evolution of an initial state corresponding to immersing a single impurity into the ground state of a  homogeneous BEC. We implement a many body generalization of the Weisskopf-Wigner method, widely used in quantum optics to study the interaction of few level atoms with the electromagnetic radiation field\cite{qoptics}. This method provides a resummation of the perturbative series in \emph{real time} and offers an alternative framework to study non-equilibrium dynamics of many body systems.

We start with a review of this framework which, to the best of our knowledge, has not been applied to the study of non-equilibrium dynamics of polaron formation.

Consider a system whose total Hamiltonian is given by $H=H_0+H_I$,  where $H_0$ describes free particles and $H_I$ is the interaction between the different degrees of freedom.    The time evolution of states in the interaction picture
of $H_0$ is given by
\be i \frac{d}{dt}|\Psi(t)\rangle_I  = H_I(t)\,|\Psi(t)\rangle_I,  \label{intpic}\ee
where the interaction Hamiltonian in the interaction picture is
\be H_I(t) = e^{iH_0\,t} H_I e^{-iH_0\,t} \,. \label{HIoft}\ee

Equation (\ref{intpic}) has the formal solution
\be |\Psi(t)\rangle_I = U(t,t_0) |\Psi(t_0)\rangle_I \label{sol}\ee
where
\be U(t,t_0)= e^{iH_0t}\,e^{-iH(t-t_0)}\,e^{-iH_0t_0}\,, \label{Utto} \ee  is  the time evolution operator in the interaction picture, it    obeys \be i \frac{d}{dt}\,U(t,t_0)  = H_I(t)U(t,t_0)~~;~~ U(t_0,t_0)=1\,. \label{Ut}\ee

   We will assume that $\langle n|H_I|n\rangle =0$ by redefining the non-interacting Hamiltonian $H_0$ to include any possible  diagonal matrix elements of the interaction. This amounts to diagonalizing the perturbation to first order in the interaction and subtracting  from $H_I$ the diagonal matrix elements in the basis $\ket{n}$.

 We   expand the time evolved state in  the basis of eigenstates of $H_0$, namely \be |\Psi(t)\rangle_I = \sum_n \ma_n(t) |n\rangle \label{decom}\ee where $|n\rangle$ obeying $H_0|n\rangle = E_n |n\rangle $ form a complete set of orthonormal states and $\ma_n(t)$ are the corresponding time dependent amplitudes. In the many body  case these are  many-particle Fock eigenstates of $H_0$. From eq.(\ref{intpic}) one finds the {\em exact} equation of motion for the coefficients $\ma_n(t)$, namely

\be \dot{\ma}_n(t) = -i \sum_m \ma_m(t) \langle n|H_I(t)|m\rangle \,. \label{eofm}\ee

\begin{figure}[ht!]
\begin{center}
\includegraphics[height=3in,width=3in,keepaspectratio=true]{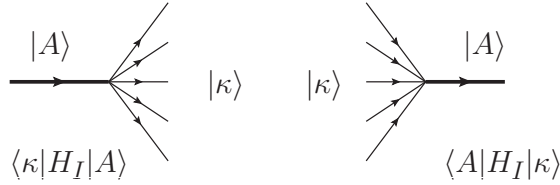}
\caption{Transitions $|A\rangle \leftrightarrow |\kappa\rangle$ in first order in $H_I$.}
\label{fig1:coupling}
\end{center}
\end{figure}

Although this equation is exact, it generates an infinite hierarchy of simultaneous equations when the Hilbert space of states spanned by $\{|n\rangle\}$ is infinite dimensional. However, this hierarchy can be truncated by considering the transition between states connected by the interaction Hamiltonian at a given order in $H_I$. Thus
consider the situation depicted in Figure~\ref{fig1:coupling} where one state, $|A\rangle$, couples to a set of states $\left\{|\kappa\rangle\right\}$, which couple back   to $|A\rangle$ via $H_I$.

Keeping only these transitions, we obtain \bea \dot{\ma}_A(t) & = & -i \sum_{\kappa} \langle A|H_I(t)|\kappa\rangle \,\ma_\kappa(t)\label{CA}\\
\dot{\ma}_{\kappa}(t) & = & -i \, \ma_A(t) \langle \kappa|H_I(t) |A\rangle \label{Ckapas}\eea where the sum over $\kappa$ is over all the intermediate states coupled to $|A\rangle$ via $H_I$.

More explicitly, for this study the state $|A\rangle \equiv \ket{1^i_{\vp};0^B}$ is the state with one impurity of momentum $\vp$ and the vacuum of the (BEC), and the states $|\kappa\rangle = \ket{1^i_{\vp-\vk};1^B_{\vk}}$ are the states with the impurity with momentum $\vp-\vk$ and a phonon of momentum $\vk$ (see below). These states are connected via the interaction Hamiltonian $H_I$ (\ref{HI}). The approximation of considering the states connected to the initial state in lowest order in perturbation theory effectively provides a resummation of \emph{second order} self energy diagrams as shown explicitly below.

Consider the initial value problem in which at time $t=0$ the state of the system $|\Psi(t=0)\rangle = |A\rangle$, i.e., \be \ma_A(0)= 1,\   \ma_{\kappa}(0) =0 .\label{initial}\ee  Solving eq.(\ref{Ckapas}) with the initial condition (\ref{initial}), and inserting its solution back into  eq.(\ref{CA}) we find \bea  \ma_{\kappa}(t) & = &  -i \,\int_0^t \langle \kappa |H_I(t')|A\rangle \,\ma_A(t')\,dt' \,,\label{Ckapasol}\\ \dot{\ma}_A(t) & = & - \int^t_0 \Sigma(t,t') \, \ma_A(t')\,dt' \,,\label{intdiff} \eea where the self-energy $\Sigma(t,t')$ is given by \be \Sigma(t,t') = \sum_\kappa \langle A|H_I(t)|\kappa\rangle \langle \kappa|H_I(t')|A\rangle \,. \label{sigma} \ee    This integro-differential equation  with {\em memory} yields a non-perturbative solution for the time evolution of the amplitudes and probabilities. Inserting the solution for $\ma_A(t)$ into eq.(\ref{Ckapasol}) one obtains the time evolution of amplitudes $\ma_{\kappa}(t)$ from which we can compute  the time dependent probability to populate the state  $|\kappa\rangle$, namely, $|\ma_\kappa(t)|^2$. This is the essence of the Weisskopf-Wigner\cite{ww} non-perturbative method ubiquitous in quantum optics\cite{qoptics}.

The hermiticity of the interaction Hamiltonian $H_I$, together with the initial conditions in eqs.(\ref{initial}) yields the unitarity condition (see appendix (\ref{app:unitarity}) for a proof.)

\be \sum_n |\ma_n(t)|^2 =1\,, \label{unitarity1}\ee where the sum is over all states. This condition will   be relevant in the discussion of the dressing dynamics  of the asymptotic state, and the entanglement entropy.

 \subsection{\label{sec:exact} Exact Solution of eqn. (\ref{intdiff}).}

  Using Eqn.(\ref{HIoft}) and the expansion in the basis of eigenstates of $H_0$ in the matrix elements of  eqs.(\ref{CA}) and (\ref{Ckapas}), we find
\be \ma_\kappa(t) = -i \langle \kappa|H_I|A\rangle\,\int^t_0 e^{i(E_\kappa - E_A)t'}\,\ma_A(t')\,, \label{Ckapaminko}\ee and \be \Sigma(t-t') = \sum_\kappa |\langle A|H_I|\kappa\rangle|^2\,e^{i(E_A-E_{\kappa})(t-t')} \equiv \int_{-\infty}^\infty d\omega'\,\rho(\omega')\, e^{i(E_A-\omega')(t-t')}\,, \label{sigmaminko} \ee where we introduced the spectral density $\rho(\omega')$,  given by \be \rho(\omega') = \sum_\kappa |\langle A|H_I|\kappa\rangle|^2 \delta(E_\kappa-\omega')\,. \label{specdens}\ee Now the integro-differential equation (\ref{intdiff}) can be solved via Laplace transform.  Introducing the Laplace variable $s$, the Laplace transform of the self-energy $\Sigma(t-t')$ is given by
\be \widetilde{\Sigma}(s) = \int_{-\infty}^\infty d\omega' ~ \frac{\rho(\omega')}{s+i(\omega'-E_A)} \label{laplasigma}\,. \ee

Defining the Laplace transform of $\ma_A(t)$ as $\mathcal{C}_A(s)$, with the initial condition $\ma_A(t=0)=1$, we find \be \mathcal{C}_A(s)= \frac{1}{ s+\widetilde{\Sigma}(s) } \,.\label{Lapla} \ee

This expression makes explicit that the Wigner-Weisskopf approximation is akin to a Dyson (geometric) resummation of self-energy diagrams similar to the Dyson series for single particle Green's function.

The  solution for the amplitude is given by the anti-Laplace transform, namely  \be \ma_A(t) = \int^{i\infty+\epsilon}_{-i\infty +\epsilon} \frac{ds}{2\pi\,i} ~\mathcal{C}_A(s)\,e^{st} \label{invlapla}\ee where   $\epsilon \rightarrow 0^+$ determines the Bromwich contour in the complex $s$-plane parallel to the imaginary axis to the right of all the singularities, for which  stability requires their real part to be negative. Writing $s=i(\omega-i\epsilon)$ we find \be \ma_A(t) = \int_{-\infty}^{\infty} \frac{d\omega}{2\pi\,i}~ \frac{e^{i\omega t}}{\Bigg[\omega-i\epsilon - \int_{-\infty}^{\infty} d\omega'~\frac{\rho(\omega')}{\omega+\omega'-E_A-i\epsilon} \Bigg]  \,. }\label{CAfin}\ee

In the free case where $\rho =0$, the pole is located at $\omega =i\epsilon \rightarrow 0$, leading to a constant $\ma_A(t)=1$. In perturbation theory, for weak coupling,  there is a complex pole very near $\omega =0$ which can be obtained directly by
expanding the integral in the denominator near $\omega =0$. We find \be  \int_{-\infty}^{\infty} d\omega'~\frac{\rho(\omega')}{\omega+\omega'-E_A-i\epsilon} \simeq -\Delta E_A - z_A\,\omega + i \,\frac{\Gamma_A}{2} \label{aproxi}\ee where \bea \Delta E_A  & = & \mathcal{P} \int_{-\infty}^{\infty} d\omega' \, \frac{\rho(\omega')}{(E_A-\omega')} \equiv {\sum_\kappa}^{'} \frac{|\langle A|H_I|\kappa\rangle|^2} {E_A-E_\kappa} \label{energyshift} \\ \Gamma_A & = & 2\pi\,\rho(E_A) \label{width} \\ z_A & = & \mathcal{P} \int_{-\infty}^{\infty} d\omega' \, \frac{\rho(\omega')}{(E_A-\omega')^2}\equiv {\sum_\kappa}^{'} \frac{|\langle A|H_I|\kappa\rangle|^2} {(E_A-E_\kappa)^2}\label{smallz}\eea where $\mathcal{P}$ stands for the principal part, and the ${\sum_\kappa}^{'} $ only sums states with $E_\kappa \neq E_A$. The term $\Delta E_A$ is recognized as the energy renormalization (Lamb shift) while $\Gamma_A$ is seen to be the decay rate as obtained from Fermi's golden rule. The {\em long time} limit of $\ma_A(t)$ is determined by this complex pole near the origin leading to the asymptotic behavior \be \ma_A(t)\simeq \mathcal{Z}_A \, e^{-i\Delta E^r_A\,t}\,e^{-\frac{\Gamma^r_A}{2}\,t} \label{tasi}\ee where
\be \mathcal{Z}_A = \frac{1}{1+z_A}\simeq 1-z_A = \frac{\partial}{\partial E_A} \big[ E_A + \Delta E_A \big] \label{wavefunc}\ee is the wave function renormalization constant (residue), and
\bea \Delta E^r_A & = &  \mathcal{Z}_A\,\Delta E_A \,, \label{DeltaEr}\\ \Gamma^r_A & = &  \mathcal{Z}_A\,\Gamma_A \,.\label{GammaAr} \eea

\subsection{\label{subsec:markov} Markovian approximation}

The time evolution of $\ma_A(t)$ determined by eq.(\ref{intdiff}) is \emph{slow} in the sense that
the time scale is determined by a weak coupling kernel $\Sigma \propto H^2_I$. This allows us to use a \emph{Markovian} approximation in terms of a
consistent expansion in derivatives of $\ma_A$. Define \be W_0(t,t') = \int^{t'}_0 \Sigma(t,t'') \, dt'' \label{Wo}\ee so that \be \Sigma(t,t') = \frac{d}{dt'}W_0(t,t'),\quad W_0(t,0)=0. \label{rela} \ee Integrating by parts in eq.(\ref{intdiff}) we obtain \be \int_0^t \Sigma(t,t')\,\ma_A(t')\, dt' = W_0(t,t)\,\ma_A(t) - \int_0^t W_0(t,t')\, \frac{d}{dt'}\ma_A(t') \,dt'. \label{marko1}\ee The second term on the right hand side is formally of \emph{fourth order} in $H_I$ because $W_0 \simeq H^2_I$ and $ \dot{\ma}_A \simeq \Sigma \simeq H^2_I$. This procedure can be iterated,  setting \be W_1(t,t') = \int^{t'}_0 W_0(t,t'') \,  dt'', \quad W_1(t,0) =0 \,,\label{marko2} \ee with
 \be W_0(t,t') = \frac{d}{dt'} W_1(t,t')\,.\label{derW1} \ee

 Integrating by parts again yields  \be \int_0^t W_0(t,t')\, \frac{d}{dt'}\ma_A(t') \,dt' = W_1(t,t)\,\dot{\ma}_A(t) +\cdots \label{marko3} \ee leading to   \be \int_0^t \Sigma(t,t')\,\ma_A(t')\, dt' = W_0(t,t)\,\ma_A(t) - W_1(t,t)\,\dot{\ma}_A(t) +\cdots \,. \label{histoint}\ee The integro-differential equation (\ref{intdiff}) now becomes
 \be \dot{\ma}_A(t) \left[1- W_1(t,t)\right] + W_0(t,t)\ma_A(t) =0 \label{markovian}\ee with the result \be \ma_A(t) = \ma_A(0)\, e^{-i\int_0^t \mathcal{E}(t')dt'}\,,\label{ampoft} \ee where \be  \mathcal{E}(t) = \frac{-i\,W_0(t,t)}{1-W_1(t,t)} \simeq -i\,W_0(t,t)\left[1+W_1(t,t)+\cdots\right]\,, \label{solumarkov}\ee with  $W_0 \simeq H^2_I~;~ W_0  W_1 \simeq H^4_I$ etc. The leading order solution of the Markovian approximation is obtained by keeping $\mathcal{E}(t) = -i\, W_0(t,t)$;  this is the order that we will consider in this study.

  Note that in general $\mathcal{E}(t)$ is complex in which the real part of $\mathcal{E}$ yields a time dependent phase while its imaginary part   determines a   \emph{ time dependent decay function}.  As discussed below in more detail, the contributions to this decay function that \emph{do not grow in time} at asymptotically long time  yield the overall asymptotic normalization of the state, namely the wave-function renormalization constant.

In the Markovian approximation the amplitudes $\ma_\kappa(t)$ become \be \ma_{\kappa}(t)   =  -i\,  \ma_A(0)\, \langle \kappa |H_I|A\rangle \,\int_0^t   e^{-i(E_A-E_\kappa)t'} \, e^{-i\int_0^{t'} \mathcal{E}(t'')\,dt''}\,dt' \,. \label{Ckapasolmarkov} \ee Therefore, once we find $\mathcal{E}(t)$ we can obtain the amplitudes of the excited states in the total wavefunction.

  With $\Sigma(t,t')$ given by eq.(\ref{sigmaminko}) to leading order in $H_I$,  we find
\be  \,\mathcal{E}(t) = -i\int^t_0 \Sigma(t,t')\,dt' = -iW_0(t,t) =  \int_{-\infty}^{\infty} d\omega' \, \frac{\rho(\omega')}{( E_A-\omega')}\,\Bigg[ 1-e^{-i(\omega'-E_A)t} \Bigg]\,,\label{Wominko} \ee  yielding  \bea -i\int^t_0 \mathcal{E}(t')\,dt' & = &
-i\, t\,\int_{-\infty}^{\infty} d\omega' \, \frac{\rho(\omega')}{( E_A-\omega')}\,\Bigg[ 1-\frac{\sin(\omega'-E_A)t}{(\omega'-E_A)t} \Bigg] \nonumber \\ & - &    \int_{-\infty}^{\infty} d\omega' \, \frac{\rho(\omega')}{( E_A-\omega')^2}\,\Bigg[ 1-\cos\big[(\omega'-E_A)t\big] \Bigg] \,. \label{energyminko}\eea

Asymptotically as $t\rightarrow \infty$, these integrals approach:
 \be \int_{-\infty}^{\infty} d\omega' \, \frac{\rho(\omega')}{( E_A-\omega')}\,\Bigg[ 1-\frac{\sin(\omega'-E_A)\,t}{(\omega'-E_A)\,t} \Bigg] ~~\overrightarrow{t\rightarrow \infty} ~~ \mathcal{P}\int_{-\infty}^{\infty} d\omega' \, \frac{\rho(\omega')}{( E_A-\omega')} \,, \label{realpartofE}\ee and
 \be \int_{-\infty}^{\infty} d\omega' \, \frac{\rho(\omega')}{( E_A-\omega')^2}\,\Bigg[ 1-\cos\big[(\omega'-E_A)t\big] \Bigg]~~ \overrightarrow{t\rightarrow \infty} ~~\pi\,t\, \rho(E_A) + \mathcal{P} \int_{-\infty}^{\infty} d\omega' \, \frac{\rho(\omega')}{( E_A-\omega')^2} \,. \label{imagE}\ee Using these results we find  in the asymptotic late time limit,

 \be -i \int^{t}_0 \mathcal{E}(t')\,dt' \rightarrow -i \Delta E_A~t - \frac{\Gamma_A}{2}~t -z_A \,, \label{phaseminko} \ee where $\Delta E_A,\Gamma_A,z_A$ are given by eqs. (\ref{energyshift},\ref{width},\ref{smallz}) to leading order in the interaction strength. From this we read off \be \ma_A(t) = \mathcal{Z}_A\,e^{-i\Delta E_A~t}\,e^{-\frac{\Gamma_A}{2}~t} ~~;~~ \mathcal{Z}_A =  e^{-z_A}\,.  \label{CAmarkovres}\ee

 Hence, the Weisskopf-Wigner framework within the Markov approximation yields results in complete agreement with the   asymptotic result from the exact solution eq.(\ref{tasi}), to leading order in $H_I$ ($  \mathcal{O}(H^2_I)$).

    However, extracting  the time evolution during the early and intermediate regimes, not just the asymptotics,   from the Laplace transform method is  more   difficult because in general, the integrals in Eqn. (\ref{CAfin}) must  be done numerically.

  The main advantage of the Weisskopf-Wigner method is that it provides a systematic framework to study the full time   evolution. Once   the spectral density $\rho(\omega)$ is obtained the final expression (\ref{ampoft}) along with the result (\ref{energyminko})  are amenable to   analytic study.

\section{Polaron dynamics.}\label{sec:polaron}

We now implement this  method   to study the time evolution of the quantum state corresponding to an impurity immersed, or ``quenched'' at time $t=0$  into the zero temperature condensate in its ground state. We identify the state $\ket{A}$ in the previous section with the state $\ket{1^i_{\vp};0^B}$ namely one impurity of momentum $\vp$ and energy $\epsilon_p=p^2/2M$ (neglecting the constant mean field energy) and the BEC vacuum,   and   the states $\ket{\kappa}$ with the excited states having one phonon, namely $\ket{1^i_{\vp-\vk};1^B_{\vk}}$. These states are connected to $\ket{1^i_{\vp};0^B}$ by the interaction Hamiltonian $H_I$ given by (\ref{HI}) .

  In the interaction picture this corresponds to
\be \ket{\Psi(0)} = \ket{1^i_{\vp};0^B}  \Rightarrow \ma^{i}_{\vp}(0)=1   \label{inistate}\ee where $\ma^{i}_{\vp}(t)$ is the amplitude of the single impurity state, and all other amplitudes vanish at the initial time. The interaction Hamiltonian $H_I$ connects the state $\ket{1^i_{\vp};0^B}$ with excited states of the form $\ket{1^i_{\vp-\vk};1^B_{\vk}}$ corresponding to an impurity with momentum $\vp-\vk$ and a phonon of momentum $\vk$,   with matrix element
\bea \bra{1^i_{\vp};0^B} H_I(t)\ket{1^i_{\vp-\vk};1^B_{\vk}} &  =  & \frac{V_k}{\sqrt{\Omega}} \,\,e^{i(\epsilon_p - \epsilon_{\vp-\vk}-E_k)t}  \label{mtxele1}\\
\bra{1^i_{\vp-\vk};1^B_{\vk}} H_I(t) \ket{1^i_{\vp};0^B}  & = &  \frac{V_k}{\sqrt{\Omega}} \, \,e^{-i(\epsilon_p - \epsilon_{\vp-\vk}-E_k)t}\,,\label{mtxele2}\eea where $V_k$ is given by Eqn. (\ref{Vk}). The matrix elements (\ref{mtxele1},\ref{mtxele2}) describe  the \emph{absorption} and \emph{emission} respectively of a phonon of momentum $\vk$ by the impurity.

Introducing $\ma^{iB}_{\vp,\vk}(t)$ as the time dependent amplitudes of the excited states $\ket{1^i_{\vp-\vk};1^B_{\vk}}$ in the total wave function $\ket{\Psi(t)}_I$ in the interaction picture,  the Weisskopf-Wigner equations (\ref{CA},\ref{Ckapas}) become
\bea \dot{ \ma}^{i}_{\vp}(t) & = & -i\,\sum_{\vk} \bra{1^i_{\vp};0^B} H_I(t)\ket{1^i_{\vp-\vk};1^B_{\vk}}\,\,\ma^{iB}_{\vp,\vk}(t) ~~;~~ \ma^{i}_{\vp}(0)=1\,,\label{ampimp} \\
\dot{ \ma }^{iB}_{\vp,\vk}(t) & = & -i\,\bra{1^i_{\vp-\vk};1^B_{\vk}} H_I(t) \ket{1^i_{\vp};0^B} \,\,\ma^{i}_{\vp}(t) ~~;~~\ma^{iB}_{\vp,\vk}(0)=0 \,. \label{ampphon}\eea The solution of  equation (\ref{ampphon}) is given by
\bea \ma^{iB}_{\vp,\vk}(t) & = & -i\, \int^t_0 \bra{1^i_{\vp-\vk};1^B_{\vk}} H_I(t') \ket{1^i_{\vp};0^B} \,\,\ma^{i}_{\vp}(t')\,dt'   \nonumber \\ & = & -i\, \bra{1^i_{\vp-\vk};1^B_{\vk}} H_I(0) \ket{1^i_{\vp};0^B}\,  \int^t_0 \,e^{-i(\epsilon_p-\epsilon_{\vp-\vk}-E_k)t'}\,\,\ma^{i}_{\vp}(t')\,dt'\,, \label{solAiB}\eea which upon inserting into (\ref{ampimp}) yields
\be \dot{ \ma}^{i}_{\vp}(t) = - \int^t_0 \Sigma(t-t')\,   \ma^{i}_{\vp}(t')\,dt' ~~;~~\ma^{i}_{\vp}(0)=1\,, \label{ampwweq}\ee where the self energy is given by
\be \Sigma(t-t') = \sum_{\vk} |\bra{1^i_{\vp-\vk};1^B_{\vk}} H_I(0) \ket{1^i_{\vp};0^B}|^2\,\,e^{i(\epsilon_p-\epsilon_{\vp-\vk}-E_k)(t-t')} \equiv \int dp_0 \, \rho(p_0;p)\,e^{i(\epsilon_p-p_0)(t-t')}\,.\label{SErho}\ee
where we introduced the spectral density
\be \rho(p_0;\vp) = \sum_{\vk} |\bra{1^i_{\vp-\vk};1^B_{\vk}} H_I(0) \ket{1^i_{\vp};0^B}|^2\,\,\delta(p_0-\epsilon_{\vp-\vk}-E_k)\,.   \label{rho} \ee

\begin{figure}[ht!]
\begin{center}
\includegraphics[height=3in,width=3in,keepaspectratio=true]{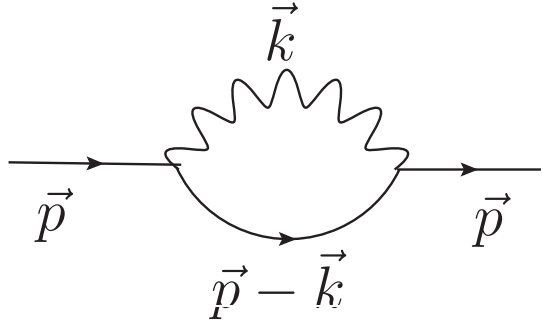}
\caption{One loop self energy $\Sigma$, the solid lines correspond to the impurity, the wavy line to phonons.}
\label{fig:selfenergy}
\end{center}
\end{figure}

We recognize $\Sigma$ as  the  one-loop self energy depicted in fig.(\ref{fig:selfenergy}). To this (second) order, the Weisskopf-Wigner method provides a real-time resummation of these self energy diagrams.

In terms of the coupling $\lambda$ given by Eqn. (\ref{lambda}) it follows that

\be \Big|\bra{1^i_{\vp-\vk};1^B_{\vk}} H_I(0)\ket{1^i_{\vp};0^B}\Big|^2 = \frac{(2\pi)^2 \lambda^2 }{\Omega} \, \frac{k}{\sqrt{1+\frac{k^2}{k^{*\,2}}}}\,,  \label{matel}\ee and the spectral density is given by
\be \rho(p_0;p)= \lambda^2\,\int^\infty_0 \frac{k^3}{\sqrt{1+\frac{k^2}{k^{*2}}}}\,\int^1_{-1}d(\cos(\theta))\,\delta\Big(\widetilde{p}_0-\frac{k^2}{2M}+
v\,k\,cos(\theta)-E_k\Big)\, dk ~~;~~ v = \frac{p}{M} \,, \label{rhoexp}\ee where $v$ is the velocity of the impurity, and we have made explicit that the spectral density depends on the variable
\be \widetilde{p}_0 = p_0 -\epsilon_p \,.\label{wpo} \ee

The explicit calculation of the spectral density is relegated to appendix  (\ref{app:specdens}), and given by Eqn. (\ref{rhofina}). For $\beta \simeq 1$ and for $\wpo \ll Mc^2/2$ it is given by Eqn. (\ref{rhosmallsapp}) (see also next section).

Following the steps described in the previous section, we find, in the Markov approximation
\be \ma^i_{\vp}(t) = e^{-i\Delta \epsilon_p(t)\,t} \, e^{-\gamma_p(t)}\,, \label{marksolA}\ee where $\Delta \epsilon_p(t)$ is the (time dependent)  energy renormalization (Lamb shift), given by (see equation  (\ref{energyminko}))
\be \Delta \epsilon_p(t) = -\int^\infty_{-\infty} d\wpo \, \frac{\rho(\wpo,p)}{\wpo}\,\Big[1-\frac{\sin(\wpo\,t)}{\wpo\,t} \Big] \,,\label{deleps}\ee
and
\be \gamma_p(t) = \int^\infty_{-\infty} d\wpo \, \frac{\rho(\wpo,p)}{\wpo^{\,2}} \,\Big[1-\cos(\wpo\,t) \Big]\,,  \label{lilgama}\ee which describes the   relaxation and dressing dynamics of the polaron.  We note that $\gamma_p(t)$ is manifestly \emph{positive}. Since our focus is to study the dynamics of relaxation and dressing,   we will neglect the energy renormalization $\Delta \epsilon_p$,  and focus solely on $\gamma_p(t)$.

Keeping only the one-phonon processes to leading order in the interaction, as described above, the time evolved state in the interaction picture is given by (we suppress the interaction picture label $I$, to simplify notation)
\be \ket{\Psi(t)}  = \ma^i_{\vp}(t)\,\ket{1^i_{\vp};0^B} + \sum_{\vk} \ma^{iB}_{\vp,\vk}(t) \ket{1^i_{\vp-\vk};1^B_{\vk}} \,. \label{entanstate}\ee This is an \emph{entangled state} of the impurity and excited phonons in the BEC revealing \emph{correlations} between the impurity and the bath of excitations of the BEC.

Once we obtain the amplitude $\ma^i_{\vp}(t)$ given by Eqn. (\ref{marksolA})  in the Markov approximation, we insert this result into Eqn. (\ref{solAiB}) to obtain the amplitudes $\ma^{iB}_{\vp,\vk}(t)$, thereby obtaining the full quantum state to this order. We obtain
\be \ma^{iB}_{\vp,\vk}(t) =  \bra{1^i_{\vp-\vk};1^B_{\vk}} H_I(0) \ket{1^i_{\vp};0^B}\,\,   \int^t_0 \,e^{-i\big(\epsilon_p+\Delta \epsilon_p(t')-\epsilon_{\vp-\vk}-E_k\big)t'}\,\,e^{-\gamma_p(t')} \,dt' \,,\label{ampexs} \ee  where the matrix element is given by (\ref{mtxele2}).

The interaction picture state (\ref{entanstate}) is remarkably similar to the \emph{variational state} for the Bose polaron proposed in ref.\cite{sarma}, which is a generalization of a variational ansatz for the Fermi polaron introduced in ref.\cite{chevy}. However there are two main differences with the variational approach of ref.\cite{sarma}:

\vspace{1mm}

\textbf{i:)} the coefficients $ \ma^i_{\vp}(t); \ma^{iB}_{\vp,\vk}(t)$ depend on time and are completely determined by the time evolution of the initial state via the Weisskopf-Wigner equations, \emph{not} from a variational approach.

\textbf{ii:)} unitary time evolution entails that these coefficients satisfy the unitarity condition $|\ma^i_{\vp}(t)|^2+ \sum_{\vk} |\ma^{iB}_{\vp,\vk}(t)|^2 =1$. In particular, as discussed below, for the case $\beta >1$  the coefficient of the single impurity term vanishes in the long time limit.

\vspace{1mm}

\textbf{Interpretation of amplitudes:}

The Weisskopf-Wigner amplitudes $\ma^i_{\vp}(t); \ma^{iB}_{\vp,\vk}(t) $ have important physical interpretations. The single impurity amplitude in the interaction picture  is given by (see Eqns. (\ref{sol}),(\ref{Utto}))
\be \ma^i_{\vp}(t) =  <{\Psi(0)}|{\Psi(t)}> = \bra{\Psi(0)}e^{iH_0t}\,e^{-iHt}  \ket{\Psi(0)}\,. \label{fide1} \ee The \emph{survival} probability
\be \mathcal{P}(t) = |<{\Psi(0)}|{\Psi(t)}>|^2 = |\ma^i_{\vp}(t)|^2 = e^{-2\gamma_p(t)} \,, \label{fidelity} \ee is  recognized as the Loschmidt echo\cite{pasta} for an initial state that evolves under a perturbation $H_I$ and is also identified with the \emph{fidelity}\cite{fidelity} of the state, which is a benchmark for quantum information\cite{qinfo}.

The \emph{probability} associated with the excited phonon states  in the time evolved state (\ref{entanstate}),   $| \ma^{iB}_{\vp,\vk}(t) |^2$, also has an illuminating interpretation: it is  the \emph{phonon} distribution function in the   time-evolved state, namely
\be f_{\vk}(t) = \bra{\Psi(t)}b^\dagger_{\vk} b_{\vk}  \ket{\Psi(t)} = | \ma^{iB}_{\vp,\vk}(t) |^2 \,. \label{phonondist} \ee Therefore the total number of phonons excited by the non-equilibrium dynamics of the impurity is given by
\be N^{ph}(t) = \sum_{\vk} | \ma^{iB}_{\vp,\vk}(t) |^2  \rightarrow \Omega\,\int \frac{d^3k}{(2\pi)^3}\,f_{\vk}(t)\,,  \label{totphonon}\ee where in the last step we took the large volume limit. This observation will become relevant when we discuss unitarity of the time evolution and the entanglement entropy in the following sections.

\subsection{Early time dynamics:} If there is a maximum momentum or frequency in the phonon spectrum, namely a cutoff $\widetilde{\Lambda}_0$ in the spectral density so that $\rho(\wpo) =0$ for  $\wpo \geq  \widetilde{\Lambda}_0$,  then for short times so that $t\widetilde{\Lambda}_0 \ll 1$ we can replace $1-\cos(\wpo\,t) \simeq \frac{{\wpo}^{\,2}}{2}\,t^2 $ in Eqn. (\ref{lilgama}). In this case we find
\be \gamma_p(t) = \frac{1}{2} \Big( \frac{t}{t_Z}\Big)^2 \,,\label{zenogama}\ee which is a manifestation of the \emph{quantum zeno} effect\cite{misra,naka,review,zenopasca}. The probability of remaining in the initial state, namely the fidelity,  at short time is given by
\be |\ma^i_{\vp}(t)|^2 = e^{-(t/t_Z)^2}\,,  \label{zenoproba}\ee
 with the \emph{Zeno} time scale $t_Z$ given by
\be \frac{1}{t^2_Z} = \int   \rho(\wpo;p)\,  d\wpo = \bra{1^i_{\vp};0^B}H^2_I \ket{1^i_{\vp};0^B} \,. \label{tZeno}\ee We used the definition of the spectral density given by Eqn. (\ref{rho}) along with the completeness of the states $\ket{1^i_{\vp-\vk};1^B_{\vk}}$. Using the result (\ref{largeporho}) in appendix (\ref{app:specdens}) for the large $\wpo$ limit of the spectral density we find
\be t_Z \simeq \Big[  \frac{2}{3} \lambda^2 k^* (2\mu)^{3/2}\,\widetilde{\Lambda}^{\,3/2}_0 \Big]^{-1/2} \,, \label{tzlam}\ee where $\mu = mM/(m+M)$ is the reduced mass.

The quantum Zeno effect has been observed in transitions of   hyperfine states of $^{9}Be$ ions in a Penning trap\cite{zenoobs1}, as well as in trapped cold $Na$ atoms\cite{zenoobs2}. In ref.\cite{zenopolaron} the authors propose that off-resonant Raman scattering off a BEC can produce a polaron state, and that optical monitoring of its decay into BEC excitations can display the quantum Zeno effect. However, the analysis of ref.\cite{nielsen} suggests that such an effect would occur on an  experimentally unobservable time scale.

For $t\widetilde{\Lambda}_0 \gg 1$ but still for short time scales with $1/\widetilde{\Lambda}_0 \ll t \ll 1/mc^2, 1/Mc^2$, we can take the frequency cutoff to infinity and change variables to  $ x = \wpo\,t  $ in (\ref{lilgama}) to obtain
\be \gamma_p(t) = t\,\int^\infty_{-\infty} \rho\big(\frac{x}{t};p\big) \,\frac{[1-\cos(x)]}{x^2}\,dx \,. \label{lilgast}\ee In this time interval  we need the behavior of $\rho(\wpo=x/t;p)$ for \emph{large} $\wpo$, which is given by (\ref{largeporho}) in appendix (\ref{app:specdens}); using this result and carrying out the remaining integral over $x$,  we find
\be \gamma_p(t) = \frac{1}{2}\sqrt{\frac{t}{t_S}} ~~;~~ t_S = \Big[16  \sqrt{\pi} \lambda^2 c\mu^{3/2}m \Big]^{-2} \,, \label{gamashortie}\ee leading to a \emph{stretched exponential} law for the fidelity, namely
\be |\ma^i_{\vp}(t)|^2 = e^{-\sqrt{\frac{t}{t_S}}}\,.  \label{shortie}\ee This result is similar to the early time dynamics for the \emph{coherence} obtained in ref.\cite{nielsen} (see discussion in section (\ref{sec:coherence})).

\subsection{Long time dynamics:}

The asymptotic long time limit  given by equation (\ref{imagE}) (with $E_A\equiv \epsilon_p$) suggests that we write
\be \rho(\wpo;p)= \rho(0;p)+\overline{\rho}(\wpo;p)\,, \label{rosplit} \ee yielding

\be \gamma_p(t) = \frac{\Gamma_p}{2}\,t + z_p(t)\,, \label{gamasplit} \ee where
\be \Gamma_p = 2\pi \rho(0;p) ~~;~~ z_p(t) =  \int^\infty_{-\infty} d\wpo \, \frac{\overline{\rho}(\wpo,p)}{\wpo^{\,2}} \,\Big[ 1-\cos(\wpo\,t)  \Big] \,,\label{asys} \ee using
\be  \int^\infty_{-\infty} d\wpo \, \,\Big[\frac{1-\cos(\wpo\,t) }{\wpo^{\,2}} \Big] = \pi\,t \,. \label{iden}\ee The result for $\Gamma_p$ is simply Fermi's Golden Rule. In the asymptotic long  time limit  the oscillatory term vanishes by the Riemann-Lebesgue theorem since $\overline{\rho}(0;p)=0$; hence it follows  from equation (\ref{imagE}) that
\be z_p(t)~~_{ \overrightarrow{t\rightarrow \infty}} ~~ z_p(\infty) =  \mathcal{P} \int^\infty_{-\infty} d\wpo \, \frac{ \rho(\wpo,p)}{\wpo^{\,2}} \,, \label{longtigam}  \ee since the principal part excises  $\wpo=0$.
Therefore, the contribution from $z_p(t)$ saturates at long time becoming a constant. This contribution describes the ``dressing'' of the impurity by the virtual cloud of phonon excitations,  whereas that from $\rho(0;p)$ yields a linear secular term indicating the decay or relaxation of the initial state via energy conserving processes. This observation leads us to separate the relaxation from the dressing dynamics in the long time limit by writing

\be e^{-\gamma_p(t)} = \underbrace{\mathcal{Z}(t)}_{~dressing~}~ \underbrace{e^{-\frac{\Gamma_p}{2}\,t} }_{~relaxation~} ~~;~~ \mathcal{Z}(t) = e^{-z_p(t)} \,.  \label{reldres}\ee

This separation entails a \emph{criterion} to distinguish between  relaxation of the \emph{quasiparticle}, namely the decay of the initial amplitude  via processes that \emph{conserve energy and momentum}  described by the decay rate $\Gamma_p$,   from the virtual transitions   that renormalize or ``dress'' the impurity by a phonon cloud,  and determine $\mathcal{Z}$. We refer to the energy conserving transitions, namely  the emission of Cerenkov phonons,  as ``on-shell'',  and the virtual transitions as ``off-shell'', with $\wpo$ describing the \emph{virtuality} of the process.

  Therefore, the survival probability, or Loschmidt echo (\ref{fidelity}) is given by
\be \mathcal{P}(t) =  {\mathcal{Z}}^2(t)~~ e^{-\Gamma_p \,t}  \,, \ee
in particular the \emph{residue}, or quasiparticle weight  is identified with $\mathcal{Z}(\infty)= e^{-z_p(\infty)}$.

In appendix (\ref{app:zinfinity}) we obtain the asymptotic value $z_p(\infty)$, (see Eqn. (\ref{zfinint})).  It is  given explicitly for the equal mass case by Eqn. (\ref{zfinir1}) below.

A rescaling $\wpo t\rightarrow x$ in the integral defining $z_p(t)$, equation (\ref{asys}) reveals that the long time dynamics is determined by the small $\wpo$ region of the subtracted spectral density. Furthermore, we focus on the case when the impurity velocity $v = p/M \simeq c$. We argue below that the long time dynamics in this region is described by the low energy Bogoliubov excitations corresponding to the linear part of the dispersion relation, namely superfluid acoustic phonons with dispersion relation   $E_k = ck$. The study of the long time dynamics in this region will yield results that are   reminiscent of \emph{critical phenomena} featuring a slowing down of relaxational and dressing dynamics (see section (\ref{sec:critical}) below). Thus our \emph{main} approximation to study the long time dynamics in this region is to restrict the dispersion relation of Bogoliubov excitations to the linear part, namely acoustic phonons with $E_k = ck$. The  regime of validity of this approximation is discussed in detail in section (\ref{sec:discussion}).

\subsection{$ v \simeq c$, small $\wpo$:}
For small $\wpo$ (long time) and $v\simeq c$ the region of support of the delta function in the spectral density (\ref{rhoexp}) corresponds to $E_k \simeq ck$. The spectral density is obtained explicitly in appendix (\ref{app:specdens}) in terms of the variables
\be \beta = \frac{v}{c}~~;~~ \epsilon_c = \frac{1}{2}\,Mc^2 ~~;~~ k_c = Mc \,.\label{newvars}\ee The ratio $\beta$ is identified with the Mach number of the impurity moving in the superfluid background of the BEC.  For $v\simeq c$ and small $\wpo$ the spectral density $\rho(\wpo)$ is given by (see Eqn. (\ref{rhosmallsapp}) in appendix (\ref{app:specdens}))

\be \rho(\wpo;p) = \frac{2\lambda^2 \, M^2\,\epsilon_c}{3 \beta  }\,\Bigg[\Big(\frac{k_{max}(\wpo)}{k_c}\Big)^3 -  \Big(\frac{k_{min}(\wpo)}{k_c}\Big)^3 \Bigg] \,, \label{rhosmalls}\ee where $k_{max}(\wpo),k_{min}(\wpo)$ are given by the following expressions  for the different cases (see appendix (\ref{app:specdens})).

For $\beta  \gtrless 1, \wpo>0 $
 \bea \frac{k_{max}(\wpo)}{k_c} & = &     (\beta-1) + \sqrt{(\beta-1)^2 + \frac{\wpo}{\epsilon_c}} \label{kmaxgb} \\
 \frac{k_{min}(\wpo)}{k_c} &  = &      -(\beta+1) + \sqrt{(\beta+1)^2 + \frac{\wpo}{\epsilon_c}}   \,, \label{kmingb} \eea

 and for $\beta >1$  and $\wpo <0$ we find

  \bea  \frac{k_{max}(\wpo)}{k_c} & = &      (\beta-1) + \sqrt{(\beta-1)^2 + \frac{\wpo}{\epsilon_c}} \label{kmaxl} \\
   \frac{k_{min}(\wpo)}{k_c} & = &      (\beta-1) - \sqrt{(\beta-1)^2 + \frac{\wpo}{\epsilon_c}} \label{kminl} \,. \eea

  For $\wpo <0$ the argument of the square roots are positive \emph{only} in the region $-P_T\leq \wpo < 0$, indicating a \emph{minimum} threshold value of $\wpo$ below which the spectral density vanishes. The  \emph{threshold} is given by
  \be P_T = \epsilon_c\,(\beta-1)^2 \,. \label{lowthre}\ee For $\beta <1$ the spectral density only has support for $\wpo \geq 0$, whereas for $\beta >1$ the spectral density also has support for $-P_T\leq \wpo < 0$. Further analysis of the emergence of this threshold is provided in appendix (\ref{app:specdens}).

 As a corollary we find the relaxation rate (\ref{asys}) in the low energy limit\footnote{It is straightforward to see that $\lambda M$ is a dimensionless parameter, which is effectively the dimensionless coupling in the low energy limit. }
 \be \Gamma_p = 2\pi\,\rho(0;p) = \frac{32\pi}{3\beta}\,(\lambda M)^2 \, \epsilon_c \,(\beta-1)^3 \, \Theta(\beta-1)\,,  \label{relarateapp}\ee where $\Theta$ is the Heavyside step function.

 For $\beta \simeq 1$  the process of   relaxation of an  impurity moving in a homogeneous condensate occurs via the Cerenkov emission of long wavelength ``on-shell'' phonons with momentum $k$ obeying the Cerenkov condition
 \be \beta\,\cos(\theta)-1 = \frac{k}{2k_c}\,,  \label{cerenkov}\ee where $\cos(\theta)=\vp\cdot \vk/pk$. For $\beta \gtrsim 1$ this condition results in   long wavelength  phonons emitted within a narrow   momentum region $ 0 \leq k \leq 2\,k_c (\beta-1)$,   within a Cerenkov ``cone'' of angular aperture $\theta_c \simeq \sqrt{\frac{2}{\beta}(\beta-1)} $.

 The power law dependence $\Gamma_p \propto (\beta-1)^3$ in (\ref{relarateapp}) is  a hallmark of the Cerenkov emission of ``on-shell'' long wavelength phonons  with   $E_k =ck$.   The smallness of the relaxation rate as $\beta \rightarrow 1^+$ is a consequence of the narrowing of the phase space available for Cerenkov phonon emission.

\section{Dressing dynamics:}\label{sec:dressing}

The  dynamical process of polaron formation is contained in $z_p(t)$, Eqn. (\ref{asys}), we now focus on obtaining its asymptotic long time limit.

 \subsection{$z_p(\infty)$}\label{sub:zpinfty}

 The explicit expression for $z_p(\infty)$ defined by Eqn. (\ref{longtigam})  is given by Eqn. (\ref{zfinint})  in
    appendix (\ref{app:zinfinity}). In the case $M=m$  it is given by
\be z_p(\infty) = \Big[\frac{2\lambda M}{\beta } \Big]^2\,\,\Bigg[1 + \frac{\beta^2-1}{2\beta} ~\ln \Bigg| \frac{1+\beta}{1-\beta}\Bigg| \Bigg] \,. \label{zfinir1} \ee Although this result has been obtained for the particular case $M=m$ and coincides with that obtained in ref.\cite{nielsen}, the cusp singularity in the derivative of $z_p(\infty)$ as $\beta \rightarrow 1$ is a \emph{general} result of the linear dispersion relation of long-wavelength phonons, as shown by the following argument.  Following the steps in appendix (\ref{app:zinfinity}), after carrying out the angular integration   and considering solely the contribution from long wavelength phononos with $k \ll k^*$, namely with   $E_k = ck$,  the integral in Eqn. (\ref{intC1}) for $z_p(\infty)$ becomes
\be z_p(\infty) = \frac{\lambda^2}{\beta c^2} \,\mathrm{Re}\Bigg\{ \int^{k^*}_0 \frac{k}{\frac{k}{2k_c}+ 1-\beta +i0^+}\,dk \, - \,  (\beta \rightarrow -\beta) \Bigg\} \simeq  (2 \lambda  M)^2   \Big[ (1-\beta) \,\ln|1-\beta| + \mathcal{C}(\beta) \Big] \,, \label{zlw} \ee where $\mathcal{C}(\beta)$ is analytic in $\beta$   and  approaches a constant as $\beta \rightarrow 1$. This analysis reveals that the \emph{singularity}   in the $\beta$ derivative of $z_p(\infty)$ as $\beta\rightarrow 1$  is solely a consequence of the linear dispersion relation of long-wavelength superfluid phonons.

 For the asymptotic time evolution three different cases arise: \textbf{i:)} $\beta < 1$,    \textbf{ii:)} $\beta = 1$, and  \textbf{iii:)} $\beta > 1$, since the spectral density is different in each case.

\subsection{$\beta < 1$}\label{sub:vlessc}

For $v<c$ the spectral density $\rho(\wpo;p)$ only has support for $\wpo >0$. As discussed above the long time limit is determined by the small $\wpo$ region of $\rho(\wpo;p)$, namely by Eqn. (\ref{rhosmalls}), with $k_{max},k_{min}$ given by eqns. (\ref{kmaxgb},\ref{kmingb}) respectively. For $\wpo \ll \epsilon_c (\beta-1)^2$ we find
\be \rho(\wpo;p) = \frac{\lambda^2}{3\beta c^4} \,\frac{\wpo^{~ 3}}{(1-\beta)^3} \Bigg[1- \Big(\frac{1-\beta}{1+\beta}\Big)^3 \Bigg] \,. \label{rholil}\ee

Because the spectral density vanishes faster than $\wpo^{~2}$ as $\wpo \rightarrow 0$, the two terms in $z_p(t)$ in Eqn. (\ref{asys}) (namely, $1$, $\cos(\wpo t)$) can be studied separately.  Hence,
\be z_p(t) = z_p(\infty) - \int^{\infty}_0 \frac{\rho(\wpo;p)}{\wpo^{~2}}\,\cos(\wpo\,t)\,d\wpo \,. \label{zsepa}\ee

The asymptotic long time limit of the second term in (\ref{zsepa}) can be obtained in a systematic asymptotic expansion in inverse powers of $t$. This is implemented by  writing
\be \cos(\wpo\,t) = \frac{1}{t} \frac{d}{d\wpo} \sin(\wpo\,t) \,, \label{deriv} \ee and integrating by parts, the ``surface term'' vanishes because $\rho \simeq \sqrt{\wpo}$ as $\wpo \rightarrow \infty$ and as $\wpo^{~3}$ as $\wpo \rightarrow 0$, hence the second term in (\ref{zsepa}) becomes
\be - \int^{\infty}_0 \frac{\rho(\wpo;p)}{\wpo^{~2}}\,\cos(\wpo\,t)\,d\wpo =  \frac{1}{t}\,\int^\infty_0  \frac{d}{d\wpo} \Bigg[\frac{\rho(\wpo;p)}{\wpo^{~2}} \Bigg]\,\sin(\wpo\,t) \,d\wpo \,.  \label{2ndstep}\ee Iterating this procedure with
\be \sin(\wpo\,t) = -\frac{1}{t} \, \frac{d}{d\wpo}\,\cos(\wpo\,t) \,,\label{3rdstep}\ee  again integrating by parts and keeping to leading order the ``surface terms'', we find
\be z_p(t)  = z_p(\infty) +  \Big( \frac{t^2_{<}}{t^2} \Big)  + \cdots \label{zlessasy}\ee
where
\be t^2_{<} = \frac{\lambda^2\,M^2}{12\,\epsilon^2_c (1-\beta)^3}\,  \Bigg[1- \Big(\frac{1-\beta}{1+\beta}\Big)^3 \Bigg]\,.\label{tless}\ee   The dots in (\ref{zlessasy}) stand for higher inverse powers of $t$. The reason that the asymptotic long time limit can be obtained as an expansion in inverse powers of $t$ is that in this case the spectral density  is an analytic function of $\wpo$ for vanishing $\wpo$.   By inspecting the region in $\wpo$ that contributes to leading order  result (\ref{zlessasy}), one finds that the physical process of dressing in this case is dominated by \emph{nearly} ``on-shell'' phonons with ``virtuality'' $\wpo = p_0 -\epsilon_p  \ll  \epsilon_c \,(1-\beta)^2$.  The   dynamical dressing time scale $t_<$ increases dramatically as $\beta \rightarrow 1^-$; this has been recognized in ref.\cite{nielsen} as a \emph{critical slowdown} of the formation dynamics of the polaron.

We emphasize that the asymptotic behavior (\ref{zlessasy}) emerges for $t \gg t_{<} \propto (1-\beta)^{-3/2}$, namely  at larger time as $\beta \rightarrow 1^{-}$.

\subsection{$\beta=1$}\label{sub:veqc}

For $\beta =1$ only $\wpo > 0$ contributes to the spectral density. The values of $k_{max},k_{min}$ are given by the $\beta \rightarrow 1$ limit of equations (\ref{kmaxgb},\ref{kmingb}) respectively, and for small $\wpo$  the spectral density is given by
\be \rho(\wpo;p) = \frac{\lambda^2 k^3_c}{3\beta\,c}\,\Big(\frac{\wpo}{\epsilon_c} \Big)^{3/2}\,\Bigg[1- \Big(\frac{\wpo}{16\,\epsilon_c} \Big)^{3/2} + \cdots \Bigg]\,,  \label{rhoveqc}\ee note that the spectral density now is \emph{non-analytic} at $\wpo =0$.
Again, both terms in $z_p(t)$ are integrable separately near the origin, upon changing variables $\wpo = x/t$, the second, time dependent term becomes
\be  -\int^\infty_{0} d\wpo \, \frac{ {\rho}(\wpo,p)}{\wpo^{\,2}} \, \cos(\wpo\,t)  = -t \, \int^\infty_{0} dx \, \frac{ \rho (x/t;p)}{x^2} \, \cos(x)\,, \label{ztveqc}  \ee which for large $t$ is dominated by the small $\wpo$ region of $\rho(\wpo;p)$. Keeping the first term in (\ref{rhoveqc}) and carrying out the remaining integral in $x$, we find for $\beta =1$
\be z_p(t) = z_p(\infty) - \Big(\frac{t^*}{t} \Big)^{1/2}+\mathcal{O}(1/t^2)~~;~~ t^* = 2\pi\,\frac{(\lambda M)^4}{\epsilon_c}\,. \label{zetapveqc}\ee The crossover between the result for $\beta<1$ with asymptotic behavior $\simeq 1/t^{2}$ to the   case with $\beta =1$ for which the asymptotic beharior is  $\simeq 1/\sqrt{t}$  occurs because the limits $\beta \rightarrow 1$ and $\wpo \rightarrow 0$  (or $t\rightarrow \infty$) of $k_{max}$, and  of $\rho(\wpo;p)$  are \emph{not uniform} and do not commute. This can be seen by comparing the small $\wpo$ limits (\ref{rholil}) for $\beta <1$ with (\ref{rhoveqc}) for $\beta =1$. The singularity for $\beta\rightarrow 1$ of (\ref{rholil}) is a consequence of taking $\wpo \ll \epsilon_c\,(1-\beta)^2$, namely, taking $\wpo \rightarrow 0$ keeping $(1-\beta)$ fixed,  whereas in (\ref{rhoveqc}) we have kept $\wpo$ fixed and taken $\beta \rightarrow 1^-$, and in this case $\rho(\wpo)$ is non-analytic for $\wpo \simeq 0$. Whereas     $\rho(\wpo)$ is \emph{analytic} in the neighborhood of $\wpo \simeq 0$ for $\beta < 1$,  for $\beta =1$ it becomes \emph{non-analytic} in this region.

\subsection{$\beta>1$}\label{sub:vlargerc}

In this case the spectral density has support both for $\wpo>0$ and $-P_T \leq \wpo \leq 0$, where the \emph{threshold} $-P_T = -\epsilon_c (\beta-1)^2$.  Because $\rho(0;p) \neq 0$ the singularity at $\wpo =0$ prevents  treating both terms in $\gamma_p(t)$ separately. Instead, we calculate the time derivative $\dot{\gamma}_p(t)$ and integrate back with the boundary condition that $\gamma_p(t) \rightarrow \frac{\Gamma_p}{2}\,t + z_p(\infty)$ as $t\rightarrow \infty$. Most of the technical details are relegated to appendix (\ref{app:ltvlarc}), the final result valid for $P_T t \gg 1$  obtained in this appendix is given by (see Eqn. (\ref{gamathreshold}) in appendix \ref{app:ltvlarc})
\be \gamma_p(t) = \frac{\Gamma_p}{2}\, t + z_p(\infty) - \Bigg[\frac{t_>}{t} \Bigg]^{3/2}\,\sin[P_T t- \pi/4] + \mathcal{O}(1/t^2)+\cdots ~~;~~ t_> = \frac{1}{\epsilon_c}\,\Bigg[\frac{ \sqrt{4\pi}\, \lambda^2 M^2}{\beta\,(\beta-1)}\Bigg]^{2/3}\,. \label{gamathreshold}\ee The $3/2$ power of time and the  power of $(\beta-1)$ in $t_>$ are distinct hallmarks of the square root singularity of the spectral density  near the threshold describing the emission and absorption of virtual phonons with virtuality $\wpo \simeq -\epsilon_c\,(\beta-1)^2$. The oscillatory part is also a consequence of the threshold\cite{fonda,maiani} and the oscillation frequency is completely determined  by $P_T = \epsilon_c\,(\beta-1)^2$. The region $\wpo \simeq 0$ of the spectral density contributes the subleading power $1/t^2$.

\section{Unitarity and  entanglement entropy.}\label{sec:info}

\subsection{Unitarity:}

As discussed in section (\ref{sec:ww}) and in appendix (\ref{app:unitarity}), the Weisskopf-Wigner method is manifestly unitary since the coefficients of the time evolved wavefunction satisfy  the condition (\ref{unitarity1}). In the case of the polaron, the coefficient $\ma^i_{\vp}(0)=1; \ma^{iB}_{\vp,\vk}(0)=0$, however, upon time evolution, the amplitude of the excited states are non-vanishing, and correlations between the impurity and phonon excitations build up as displayed by the time evolved state (\ref{entanstate}).
Therefore, unitary time evolution entails a \emph{flow} of probability from the initial state with one impurity and the BEC in the  ground state, to excited states. In this section we   study how unitarity is fulfilled. Although unitarity is an exact statement,   we have only determined the behavior of the amplitude (\ref{marksolA}), and consequently of (\ref{ampexs}) for early transient and long time. Hence a full study of the time evolution of the amplitudes  would necessarily entail an exhaustive numerical study for a large parameter and dynamical range. Instead, we will focus on understanding the fulfillment of unitarity at long time  by invoking the following approximations  valid for weak coupling.

\textbf{i:)} We neglect the time evolution of $\Delta \epsilon_p$ in (\ref{ampexs}), furthermore, we absorb this correction into a renormalization of the impurity energy: $\epsilon_p +\Delta\epsilon_p \rightarrow \epsilon^r_p$, the renormalized impurity energy (renormalized polaron mass). To simplify notation we write $\epsilon_p$ everywhere, understanding that this is the renormalized energy of the impurity.

\textbf{ii:)} We neglect the early time evolution. This is warranted because during the initial transient the amplitudes do not vary much and the $\ma^{iB}_{\vp;\vk}$ are $\propto \lambda $ and therefore small in weak coupling.

\textbf{iii:)}   Neglecting the early transient dynamics, we write $e^{-\gamma_p(t)} \equiv e^{-\frac{\Gamma_p}{2}t} \, e^{-z_p(\infty)}\,e^{-f_p(t)} $ where $f_p(t)\propto \lambda^2 M^2 t^{-\alpha}$   describes the sub-leading power laws derived above for the various cases. We will neglect this contribution because it is subleading at long time and always perturbatively small in weak coupling, keeping only the leading behavior, namely
\be  \ma^i_{\vp}(t) = e^{-\gamma_p(t)} \simeq e^{-z_p(\infty)}\,  e^{-\frac{\Gamma_p}{2}t}\,. \label{aimpapx}\ee

Implementing  these approximations we find
\be |\ma^{iB}_{\vp,\vk}(t)|^2  =     e^{-2z_p(\infty)}\,  \frac{\Big|\bra{1^i_{\vp-\vk};1^B_{\vk}} H_I(0)\ket{1^i_{\vp};0^B}\Big|^2}{\Big[\Big(\epsilon_p - \epsilon_{\vp-\vk}-E_k )  \Big)^2+ \Big(\frac{\Gamma_p}{2} \Big)^2  \Big] }\,\Big|1-e^{-i(\epsilon_p - \epsilon_{\vp-\vk}-E_k -i \frac{\Gamma_p}{2} )t} \Big|^2 \,, \label{excicoefs}    \ee    with the result (\ref{matel}) for the numerator.

Since we have neglected the early time transient and the asymptotic long time tails, both of $\mathcal{O}(\lambda^2 M^2)$, we can only consistently confirm unitarity up to $\mathcal{O}(\lambda^2 M^2)$.

Notice that when the Cerenkov condition is fulfilled, namely $\epsilon_p = \epsilon_{\vp-\vk}+E_k$, the denominator in (\ref{excicoefs}) becomes \emph{resonant}, and the width of the resonance is $\Gamma_p$.

The unitarity condition (\ref{unitarity1}) yields
\be \Big|\ma^i_{\vp}(t)\Big|^2 + \sum_{\vk}  |\ma^{iB}_{\vp,\vk}(t)|^2 =1\,.
 \label{unitarityAcoefs}\ee Writing the first term as in Eqn. (\ref{aimpapx}) and the second term   in terms of the spectral density $\rho(\wpo;p)$ (\ref{rho})  we find that the condition (\ref{unitarityAcoefs}) becomes
\be e^{-2z_p(\infty)}\,e^{-\Gamma_p t}  +  e^{-2z_p(\infty)} \, \int d\wpo \frac{\rho(\wpo;p)}{\Big[\wpo^2 + \Big(\frac{\Gamma_p}{2} \Big)^2  \Big]}\,\Big|1-e^{ i\wpo t}\,e^{-\frac{\Gamma_p}{2}\,t} \Big|^2 =1
 \label{unitarityAcoefsII}\ee

 \vspace{1mm}

 \subsubsection{$\mathbf{ \beta > 1}:$}

 \vspace{1mm}

  In this case $\Gamma_p = 2\pi \rho(0;p) \neq 0$, hence we write $\rho(\wpo;p) = \rho(0;p)+ \overline{\rho}(\wpo;p)$ in the integral in the second term in Eqn. (\ref{unitarityAcoefsII}). The contribution to the $\wpo$ integral from the term with $\rho(0;p)$ yields
 \be  \rho(0;p)\, \int d\wpo \, \frac{\Big|1-e^{ i\wpo t}\,e^{-\frac{\Gamma_p}{2}\,t} \Big|^2}{\Big[\wpo^2 + \Big(\frac{\Gamma_p}{2} \Big)^2  \Big]}\,  = 2\pi \,\frac{\rho(0;p)}{\Gamma_p}\,\Big[1-e^{-\Gamma_p t}\Big]  =  \Big[1-e^{-\Gamma_p t}\Big]\,.\label{polecont}\ee  The contribution from $\overline{\rho}(\wpo;p)$ is perturbatively small: for small $\wpo$ we find that $\overline{\rho}(\wpo;p) \simeq \lambda^2 M^2 \wpo +\cdots$, therefore the complex poles at $\wpo = \pm i\Gamma_p/2$ yield a perturbatively small contribution of $\mathcal{O}(\lambda^2M^2)$ or higher, as compared to the contribution from $\rho(0;p)$ above.  Furthermore, since $e^{-2z_p(\infty)} \simeq 1 + \mathcal{O}(\lambda^2 M^2)$ and non-secular in time we find that the left hand side of (\ref{unitarityAcoefsII}) becomes
 \be e^{-2z_p(\infty)}\,e^{-\Gamma_p t}+ e^{-2z_p(\infty)}\, \Big[1-e^{-\Gamma_p t}\Big] + \cdots = 1 +  \cdots \,,\label{unitaconf}\ee where the dots stand for high order in the coupling $\lambda^2 M^2$, thereby confirming unitarity to leading order in the coupling.  As $t\rightarrow \infty$ the contribution from the initial state $\ket{1^i_{\vp};0^B}$ vanishes and unitarity is fulfilled  by   the initial probability flowing to   excited   states of the impurity \emph{entangled} with phonons. Namely, for $t \gg 1/\Gamma_p$, to leading order in the interaction the quantum state is given by $\ket{\Psi(\infty)} = \sum_{\vk} \ma^{iB}_{\vp,\vk}(\infty) \ket{1^i_{\vp-\vk};1^B_{\vk}}$ where for $\beta \simeq 1$ the phonons in the excited states are   within the resonant band fulfilling the Cerenkov condition, namely with $0\leq k \leq 2k_c (\beta -1)$.

 An alternative manner to understand the saturation of unitarity from the excited states in the asymptotic long time is the following.  Taking the long time limit $t \gg 1/\Gamma_p$ in (\ref{unitarityAcoefsII}) and neglecting the term $e^{-2z_p(\infty)} = 1 + \mathcal{O}(\lambda^2 M^2) + \cdots $ write the integral in the second term in (\ref{unitarityAcoefsII}) as
 \be \int d\wpo \frac{\rho(\wpo;p)}{\Big[\wpo^2 + \Big(\frac{\Gamma_p}{2} \Big)^2  \Big]} = \frac{2}{\Gamma_p}\,\int d\wpo \, \rho(\wpo;p)\,\frac{\Big(\frac{\Gamma_p}{2}\Big)}{\Big[\wpo^2 + \Big(\frac{\Gamma_p}{2} \Big)^2  \Big]} \rightarrow \frac{2\pi}{\Gamma_p}\,\rho(0;p)  =1 \,,\label{asyga}\ee where for small coupling we have taken the ``narrow width limit'' ($\Gamma_p \rightarrow 0$) using $\varepsilon/ (x^2 +\varepsilon^2) \rightarrow \pi\,\delta(x) $ as $\varepsilon \rightarrow 0$. This simple argument clearly indicates that for $\beta > 1$ unitarity is (nearly) saturated by the excited states with ``on-shell'' phonons that obey the Cerenkov condition (\ref{cerenkov}),  or energy conservation, corresponding to $\wpo =0$.

\vspace{1mm}

 \subsubsection{$\mathbf{ \beta \leq 1}:$}

 \vspace{1mm}

In this case $\Gamma_p=0$ and   $e^{-2z_p(\infty)} \simeq 1- 2z_p(\infty) +\cdots$ and $ z_p(\infty) \simeq \mathcal{O}(\lambda^2 M^2)$, therefore to leading order in the coupling the left hand side of the unitarity condition (\ref{unitarityAcoefsII}) becomes
\be 1-2\,z_p(\infty) + \cdots +2\, \int d\wpo \frac{\rho(\wpo;p)}{ \wpo^2 }\,\Big[ 1- \cos(\wpo t)\Big]   = 1  \label{unitacond2}\ee The dots stand for terms of  $\mathcal{O}(\lambda^4 M^4)$ and higher. In the asymptotic long time limit the oscillatory cosine term vanishes by the Riemann-Lebesgue theorem. With the definition of $z_p(\infty)$ given by Eqn. (\ref{longtigam}) it is straightforward to see that unitarity is fulfilled up to $\mathcal{O}(\lambda^4 M^4)$.  In this case at asymptotically long time, the initial state contributes with probability $ e^{-2z_p(\infty)}$, whereas the excited states of an impurity entangled with phonons contributes $1-e^{-2z_p(\infty)} \simeq 2\,z_p(\infty)+\cdots$.

Thus we have confirmed the fulfillment of unitarity up to leading order ($\mathcal{O}(\lambda^2 M^2)$) in the interaction.

\subsection{von Neumann entanglement entropy:}

     The confirmation of the fulfillment of unitary time evolution clearly shows that the decay of the fidelity (survival probability of the initial impurity state), either by relaxation for $\beta >1$ or dressing  for $\beta \leq 1$,  results in the build up of impurity-phonon correlations
    resulting in the  entangled state (\ref{entanstate}). This flow of probability also entails a \emph{loss of information}, which can be made manifest by obtaining  the \emph{reduced density matrix} and the entanglement entropy for the impurity.

   From the time evolved state (\ref{entanstate}), the (pure state) density matrix is  $\ket{\Psi(t)}\bra{\Psi(t)}$. We obtain the impurity \emph{reduced density matrix} by tracing over the phonon degrees of freedom, namely
\be \rho^r_i(t) = \mathrm{Tr}_{BEC} \ket{\Psi(t)}\bra{\Psi(t)}  = |\ma^i_{\vp}(t)|^2 \, \ket{1^i_{\vp}}\bra{1^i_{\vp}} + \sum_{\vk} |\ma^{iB}_{\vp,\vk}(t)|^2 \, \ket{1^i_{\vp-\vk}}\bra{1^i_{\vp-\vk}}\,.
\label{rhoreduced} \ee
This reduced density matrix is diagonal in the impurity basis and describes a mixed state; the von Neumann entanglement entropy is given by
\be S(t) = - |\ma^i_{\vp}(t)|^2\,\ln|\ma^i_{\vp}(t)|^2 - \sum_{\vk}  |\ma^{iB}_{\vp,\vk}(t)|^2\,\ln  |\ma^{iB}_{\vp,\vk}(t)|^2 \,, \label{ententropy}\ee and it is a measure of the \emph{correlations} between the impurity and the excited BEC, and  of the information loss during the time evolution as the impurity relaxes and is dressed by its coupling to the phonon degrees of freedom. It is clear that  $S(0)=0$ because of the initial conditions $\ma^i_{\vp}(0)=1\,,\, \ma^{iB}_{\vp,\vk}(0)=0$, and as a consequence of unitary time evolution and the flow of probability to the excited states, $ |\ma^i_{\vp}(t)|^2 < 1\,,\, |\ma^{iB}_{\vp,\vk}(t)|^2 <1 $ for $t>0$, implying that  $S(t) >0$ for $t>0$. Therefore, the dynamics of relaxation and dressing leads to a growth of entanglement entropy as a consequence of the creation of excitations and the flow of probability from the initial to the excited states. By unitary time evolution the flow of probability to the excited states is determined by the decay of the  Loschmidt echo or fidelity (\ref{fidelity}). A similar relation between the decay of the Loschmidt echo and information flow into the environmental bath  has been shown to hold in the case of a qubit coupled to environmental degrees of freedom\cite{haikka}.

$S(t)$ is determined by the time evolution of the amplitudes,   (\ref{marksolA}, \ref{ampexs}). As in the study of unitarity above, we will focus on the long time behavior under the same approximations implemented above, with the amplitudes $\ma^i_{\vp}(t)~;~ |\ma^{iB}_{\vp,\vk}(t)|^2$, given by eqns. (\ref{aimpapx}), (\ref{excicoefs}) respectively.

\subsubsection{$\beta > 1$}\label{subsubsec:betag1}

In this case $\ma^i_{\vp}(\infty) =0$  because $\Gamma_p \neq 0$, therefore
\be S(\infty) = -   \sum_{\vk}  |\ma^{iB}_{\vp,\vk}(\infty)|^2\,\ln  |\ma^{iB}_{\vp,\vk}(\infty)|^2 \,.  \label{sfinitybetag1}\ee    In this case unitarity is saturated by the excited states that obey the Cerenkov condition (\ref{cerenkov}), and $|\ma^{iB}_{\vp,\vk}(\infty)|^2$ is largest for this (resonant) region of phonon momentum (see Eqn.(\ref{excicoefs}) ), with the maximum momentum $2k_c(\beta-1)$. Therefore we can extract the leading behavior of $S(\infty)$ by the following steps: i) in the $\ln  |\ma^{iB}_{\vp,\vk}(\infty)|^2$ keep only the momenta that obey the Cerenkov condition, resulting in the denominator in (\ref{excicoefs}) being replaced by $(\Gamma_p/2)^2$. ii) In the $\ln  |\ma^{iB}_{\vp,\vk}(\infty)|^2$ replace the momentum by the maximum value in the Cerenkov band, namely $k \rightarrow 2k_c(\beta-1)$. With these approximations for the logarithm, it follows that
\be \ln  |\ma^{iB}_{\vp,\vk}(\infty)|^2 \rightarrow \ln\Big[ \frac{\lambda^2\,k_c\,(\beta-1)}{\Omega\,\Gamma^2_p}\Big]+ \cdots \label{logaterm}\ee where the dots stand for subleading   terms in  $\lambda$ and $(\beta-1)$.  The unitarity condition in this case gives $\sum_{\vk}  |\ma^{iB}_{\vp,\vk}(\infty)|^2 =1+\cdots$ which yields the following leading order result
\be S(\infty) = \ln\Big[  \frac{\Omega\,\big(\Gamma_p/c\big)^3}{(\lambda M)^4\,(\beta-1)^4} \Big] + \cdots \,.  \label{Sfinitybetag1}\ee  The argument of the logarithm is $\gg 1$ for the following reason: consider that the volume   is a cube of side $L$ with $\Omega = L^3$. For the impurity moving with $v \simeq c$ to decay inside the volume it must be that $L >> c/\Gamma_p$ with $c/\Gamma_p$ being the decay length. Therefore $\Omega(\Gamma_p/c)^3 \gg 1$ and the smallness of the denominator in (\ref{Sfinitybetag1}) guarantees that $S(\infty)$ is positive and large. Furthermore, the dominant dynamical time scale in this case is   $1/\Gamma_p$ with subleading power laws, therefore, we conclude that for $\beta >1$ the entanglement entropy grows to its asymptotic value $S(\infty)$ on the time scale $1/\Gamma_p$, modulated with a power law multiplying an oscillatory function whose oscillation frequency is determined by the threshold at $\epsilon_c (\beta -1)^2$.

The logarithm of the volume is noteworthy. It stems from the fact that the total number of phonons produced is $\mathcal{O}(1)$: first note that the matrix element squared (\ref{matel}) is of order $1/\Omega$, this is a results of the normalization of single particle states. Secondly, equation (\ref{phonondist}) identifies the asymptotic phonon distribution function with $|\ma^{iB}_{\vp,\vk}(\infty)|^2 \propto 1/\Omega$. The total asymptotic entropy is then of the
form $\sum_{\vk} f_{\vk}(\infty) \ln[f_{\vk}(\infty)] \rightarrow \Omega \int \frac{d^3k}{(2\pi)^3} \,f_{\vk}(\infty) \ln[f_{\vk}(\infty)]$. The factor $\Omega$ cancels the $1/\Omega$ in $f_{\vk}(\infty)$ but the volume factor in the logarithm remains, this is because $\Omega \int \frac{d^3 k}{(2\pi)^3} f_{\vk}(\infty) \simeq 1$ by unitarity, namely only a number of phonons $\simeq \mathcal{O}(1)$ is created.

\subsubsection{$\beta \leq  1$}\label{subsubsec:betale1} In this case $\Gamma_p=0$,  the Cerenkov condition (\ref{cerenkov}) \emph{cannot} be fulfilled, and the denominator in  (\ref{excicoefs}) is not resonant. We extract the leading order contribution to $S(\infty)$ by the following procedure: i) in the $\ln|\ma^{iB}_{\vp,\vk}(\infty)|^2 $ rescale the momentum by $k_c$. This term becomes of the form $\ln[\lambda^2 M^2/\Omega k^3_c] + \mathcal{F}[k/k_c,\beta,r]$ with $r=m/M$. The  k-integral of $|\ma^{iB}_{\vp,\vk}(\infty)|^2\,\mathcal{F}$ is finite, with a finite limit as $\beta \rightarrow 1$ (although it may feature a cusp in its $\beta$ derivative) and of order $\lambda^2 M^2$. Using the result (\ref{unitacond2}) from the  unitarity condition, and $|\ma^i_{\vp}(\infty)| = e^{-2z_p(\infty)} \simeq 1- 2z_p(\infty) +\cdots $ we find in this case the leading order result
\be S(\infty) = 2z_p(\infty) \Big[ \ln[\Omega k^3_c/\lambda^2 M^2] + 1 \Big] +\cdots \label{sfinitybl1}\ee where the dots stand for subleading  contributions that include  the integral with $\mathcal{F}$.

For the impurity to be inside the volume $\Omega = L^3$ it must be that its de-Broglie wavelength $\propto 1/k_c  \lesssim L$, therefore  $\Omega k^3_c \gtrsim 1$. Hence, for weak coupling the logarithm in (\ref{sfinitybl1}) is $\gg 1$ justifying keeping only this contribution as the leading term in the entanglement entropy.
The origin of the volume factor in the argument of the logarithm is exactly the same as that discussed above for the case $\beta > 1$.

The time evolution of $S(t)$ approaching its asymptotic limit $S(\infty)$ is determined by the power laws obtained in the previous sections for $\beta \leq 1$.

It is noteworthy that in the non-decaying case, $\beta \leq 1$, the wave function renormalization, describing the dressing of the polaron determines the entanglement entropy. While this feature is expected since the entanglement entropy is a measure of the flow of information from the initial to the excited states, the logarithmic enhancement in terms of the volume and coupling is perhaps unexpected.

The behavior of $S(\infty)$   is strikingly different between the $\beta > 1$ and $\beta  \leq 1$ cases. The latter, given by Eqn. (\ref{sfinitybl1}) is \emph{suppressed} by $\lambda^2 M^2$ as compared to the former since $z_p(\infty) \propto \lambda^2 M^2$. Thus in the weak coupling limit, there is a sharp change in $S(\infty)$ as $\beta \rightarrow 1$ from above. This is expected on physical grounds, since for $\beta >1$ impurity relaxation leads to the creation of \emph{real} phonons, whereas for $\beta \leq 1$ the dressing of the impurity corresponds to a virtual cloud of phonons. However, while one expects a change in the behavior of the entanglement entropy, the sharp discontinuity in the weak coupling limit is perhaps not anticipated.

The identification given by (\ref{fide1}) relating the Loschmidt echo and fidelity to the survival probability, combined with the unitarity arguments presented above imply that Loschmidt echo/fidelity decay is directly related to the build up of impurity-phonon correlations and the entanglement entropy. This is a new arena in which the decay of the Loschmidt echo is directly related to decoherence and growth of entanglement entropy, previously discussed within the context of the semiclassical regime of quantum open systems\cite{paz}.

\section{Long time dynamics: a dynamical critical phenomenon?}\label{sec:critical}

Taken together the results obtained in the previous sections hint at   dynamical \emph{critical} phenomena for an impurity inmmersed in a homogeneous BEC  with Mach number $\beta \simeq 1$. The critical \emph{slow- down} of the dressing dynamics of   polaron   formation was originally anticipated in ref.\cite{nielsen}. Our study not only confirms the suggestion in this reference, but complements it in various ways that strengthen the interpretation of the long-time dynamics as a manifestation of behaviour akin to critical dynamics dominated by the low energy spectrum of superfluid phonon excitations:

\vspace{1mm}

\textbf{i):} For $\beta > 1$,  the   decay rate (\ref{relarateapp}) vanishes as $\Gamma_p \propto (\beta-1)^3$ as $\beta \rightarrow 1^+$.   This power law is a direct consequence of the behavior of the spectral density for low energy, namely long-wavelength phonons, and  it is a manifestation of the narrowing of the phase space for Cerenkov emission.  The dynamics of dressing features a  power-law $(t_>/t)^{3/2}$ with $t_> \propto (\beta-1)^{-2/3}$ modulated by an  the oscillatory function with frequency $\propto (\beta-1)^2$. This slow dressing dynamics is  a consequence of the threshold in the density of states vanishing as $(\beta -1)^2$, and again, is  determined by the long-wavelength phonon spectrum.

\vspace{1mm}

\textbf{ii):} The cusp in the $\beta$-derivative  of the  wave function renormalization, which is, indeed, a direct and \emph{universal}  consequence of the linear dispersion of long-wavelength superfluid phonons as discussed in section (\ref{sec:dressing}) (see the discussion leading to Eqn. (\ref{zlw})).

\vspace{1mm}

\textbf{iii):} A  sharp transition in the dressing dynamics  between $\beta >1$  and $\beta\leq 1$ cases: for $\beta<1$ the  dressing of the impurity, or  polaron formation  occurs asymptotically on a \emph{dynamical time scale} $t_{<} \propto (1-\beta)^{-3/2}$  with a power law $\simeq (t_</t)^2$ slowing down as $\beta \rightarrow 1^{-}$. For $\beta =1$, the power law changes to $\sqrt{t^*/t}$ with $t^*$ given by (\ref{zetapveqc}), whereas for $\beta> 1$ the dressing dynamics features a power law $(t_>/t)^{3/2}$ with $t_> \propto (\beta-1)^{-2/3}$ and is modulated by periodic oscillations with a typical frequency $\propto (\beta-1)^2$.

\vspace{1mm}

\textbf{iv):} A sharp transition at $\beta=1$ is also manifest in the asymptotic entanglement entropy. For $\beta >1$, the asymptotic value $S(\infty)$ is given by Eqn. (\ref{Sfinitybetag1}), whereas for
$\beta \leq 1$ it is given by (\ref{sfinitybl1}) where for equal mass $z_p(\infty)$ is given by (\ref{zfinir1}). For $\beta \leq 1$  the logarithms in $S(\infty)$ are multiplied by $\lambda^2 M^2\ll 1$, therefore substantially suppressed as compared to the $\beta > 1$ case. Furthermore, for $\beta > 1$  the leading contribution to $S(\infty)$ arises from real (``on-shell'') long wavelength phonons within the resonance band satisfying the Cerenkov condition resulting in a strong dependence of $S(\infty)$ on $\beta-1$, whereas for $\beta <1$, $S(\infty)$ features a finite limit as $\beta \rightarrow 1^-$.

Therefore relaxational and dressing dynamics feature a behavior very similar to critical phenomena where the Mach number $\beta$ plays a role akin to $T_c/T$, with $T$ a temperature variable and $T_c$ its critical value, in the sense that the asymptotic long time dynamics is sharply different for $\beta \lessgtr 1$. The long time dynamics of polaron formation is characterized by power laws $t^{-\alpha}$ with exponents $\alpha$ that are different for $\beta \lessgtr 1$ and pre-factors that reveal the slow-down of formation as $\beta \rightarrow 1$. The decay rate $\Gamma_p \propto (\beta-1)^3\,\Theta(\beta-1)$ is suggestive of a quantity akin to an order parameter in that it vanishes for $\beta \leq 1$ and rises with a power law for $\beta > 1$.

 The main underlying reason for the ``critical'' dynamics is the \emph{linear dispersion relation} of long wavelength superfluid phonons. This feature of the spectrum of excitations completely determines the low energy behavior of the spectral density of the impurity, hence the long time dynamics. Whereas the pre-factors of the various quantities depend on the parameters  such as $\lambda, M, c$, etc. the \emph{power laws} for dressing dynamics $t^{-\alpha}$,  the dependence $\Gamma_p \simeq (\beta-1)^3\,\Theta(\beta-1)$,  the powers $|\beta-1|^{-\xi}$ associated with the formation time scales $t_>, t_<$, the \emph{cusp}   in the $\beta$ derivative of $z_p$, and the sharp discontinuity of the asymptotic entanglement entropy at $\beta =1$  are all \emph{universal} consequences of the linear dispersion relation of long wavelength phonons.

 While all these features are strong hints of phenomena akin to  critical behavior for $\beta \simeq 1$, a deeper characterization of these non-equilibrium aspects as a novel manifestation of dynamical critical phenomena merits further study.

\section{Relation to coherence\cite{nielsen}:}\label{sec:coherence}
Many  of the results obtained above  are strikingly similar to those obtained in reference\cite{nielsen} for \emph{the coherence}. In this reference the authors consider an initial state
\be \ket{\Psi_{\theta}(0)} = \Big(\cos(\theta)+ \sin(\theta)\,C^\dagger_{\vp}\Big)\ket{0^i;0^B}\, \label{cohestate}\ee where $\ket{0^i;0^B}$ corresponds to the vacuum state both for the (BEC) and the impurity. The coherence is defined in ref. \cite{nielsen} as
\be \mathcal{G}_{\vp}(t) = \bra{\Psi_\theta(0)}C_{\vp}(t)\ket{\Psi_\theta(0)} \,, \label{coherence}\ee where $C_{\vp}(t)$ is the impurity annihilation operator in the Heisenberg picture. The authors of ref. \cite{nielsen} show that $\mathcal{G}_{\vp}(t)$ is related  to the impurity Green's function by $\mathcal{G}_{\vp}(t) =   \,  \cos(\theta)\,\sin(\theta) \bra{0^i;0^B} C_{\vp}(t) C^{\dagger}_{\vp}(0) \ket{0^i;0^B} $.

 It is definitely \emph{not obvious} that there should be any similarity between many of the results found above and those obtained in ref.\cite{nielsen} for the coherence (\ref{coherence}), since the latter \emph{vanishes} exactly for $\theta = \pi/2$ corresponding to the single impurity state that we study.

We note that the original total  Hamiltonian (\ref{totalH}) is invariant under an abelian $U(1)$ global symmetry $C_{\vp} \rightarrow e^{i\varphi}\,C_{\vp}$ with $\varphi$ a space-time constant. As a consequence,  the impurity number $N_i= \sum_{\vp}\,C^\dagger_{\vp}\, C_{\vp}$ is conserved by the dynamics.
We note, however,  that the state $\ket{\Psi_\theta(0)}$ Eqn. (\ref{cohestate}) is \emph{not} an eigenstate of $N_i$ since it is  a linear superposition of the vacuum and a one impurity state.  Therefore, it is   possible for the annihilation operator $C_{\vp}$ to acquire an expectation value in this state.

  In reference\cite{nielsen} the time evolution of the reduced density matrix for the impurity $\rho^r_i(t)$ is obtained from a quantum master equation in the Born approximation, wherein the ``bath'' (the BEC in the ground state) is traced out,  assuming a factorization between the density matrix of the bath and the reduced density matrix for the impurity at all times.  An equation of motion for the coherence is obtained from the equivalence $\mathcal{G}_{\vp}(t)= \mathrm{Tr} \, C_{\vp}(0)\,\rho^r_i(t) $.

To compare with the Weisskopf-Wigner approach of the previous sections, we write the initial state (\ref{cohestate}) as
\be \ket{\Psi_\theta(0)} = \ma^0_{\theta}(0)\ket{0^i;0^B} + \ma^i_{\vp,\theta}(0)\ket{1^i_{\vp};0^B} ~~;~~ \ma^0_{\theta}(0) = \cos(\theta)~~;~~ \ma^i_{\vp,\theta}(0)=\sin(\theta)\,.  \label{coheini}\ee Hence the initial state considered in the previous sections, (\ref{inistate}), is precisely (\ref{cohestate}) with $\theta=\pi/2$. Despite this simple relation between the initial states, there is a fundamental difference between the two:  \emph{only} for $\theta =0,\pi/2, ~\textrm{modulo} ~ 2\pi$ is the initial state (\ref{cohestate})   an eigenstate of the impurity number operator $N= \sum_{\vp} C^\dagger_{\vp}C_{\vp}$, namely has a definite impurity particle number. For any other value the initial state (\ref{cohestate}) breaks the underlying $U(1)$ global gauge invariance\footnote{It does \emph{not} transform covariantly under the $U(1)$ global phase transformation.} because it is a mixture of states with different impurity particle number.

We can now apply the Weisskopf-Wigner method to obtain the time evolved state from the initial state (\ref{cohestate}) following the steps eqns. (\ref{CA},\ref{Ckapas}) leading to the equations for the amplitudes. An important aspect in this program is that  the interaction Hamiltonian $H_I$ (\ref{HI})  is such that
\be H_I(t)  \ket{0^i;0^B} = 0 \,. \label{Hiani0}\ee As a consequence, any matrix element of the interaction Hamiltonian of the form $\bra{\alpha}H_I(t)\ket{0^i;0^B}; \bra{0^i;0^B}H_I(t)\ket{\alpha}$ vanishes for arbitary $\ket{\alpha}$, and  the amplitude of the vacuum state contribution      to the initial state (\ref{cohestate}) is independent of time, namely $\ma^0_{\theta}(t)=\ma^0_{\theta}(0)$. The reason for this is physically clear: the interaction Hamiltonian conserves the number of impurity particles, the state $\ket{0^i;0^B}$ has \emph{zero} impurities, therefore upon time evolution it must remain the impurity vacuum, and no phonon excitation can be created because such process entails the annihilation of an impurity of which there are none in the impurity vacuum. The amplitude equations for the impurity and impurity-phonon states are exactly the same as found above, namely (\ref{ampwweq},\ref{solAiB}) but now with the initial condition that $\ma^i_{\vp,\theta}(0)=\sin(\theta)$ in (\ref{ampwweq}). We find that the initial state (\ref{coheini}) evolves in time into the state
\be \ket{\Psi_\theta(t)} = \cos(\theta)\ket{0^i;0^B}+ \sin(\theta)\Bigg\{ \ma^i_{\vp}(t)\,\ket{1^i_{\vp};0^B} + \sum_{\vk} \ma^{iB}_{\vp,\vk}(t) \ket{1^i_{\vp-\vk};1^B_{\vk}} \Bigg\} \,,\label{wwcoheini}\ee
where the amplitudes $ \ma^i_{\vp}(t), \ma^{iB}_{\vp,\vk}(t)$ are precisely given by (\ref{ampwweq}, \ref{solAiB}), with  $ \ma^i_{\vp}(0)=1, \ma^{iB}_{\vp,\vk}(0)=0$. In the interaction picture $C_{\vp}(t) = e^{-i\epsilon_p t}\,C_{\vp}(0)$, therefore the coherence as defined in ref.\cite{nielsen} becomes
\be \bra{\Psi_\theta(t)}C_{\vp}(t)\ket{\Psi_\theta(t)} = \sin(\theta)\cos(\theta)\,e^{-i\epsilon_p t}\, \ma^i_{\vp}(t)\,.  \label{coherenceoft}\ee Up to a phase, and the overall multiplicative factor,  the coherence defined in ref.\cite{nielsen} is simply proportional to  the amplitude $\ma^i_{\vp}(t)$. This is the explanation of the equivalence between the results obtained via the Weisskopf-Wigner framework above and those in ref.\cite{nielsen} obtained from the quantum master equation. Furthermore, from the relation between the coherence and the impurity Green's function   established in ref.\cite{nielsen}, it follows that $e^{-i\epsilon_p t}\, \ma^i_{\vp}(t)= \bra{0^i;0^B} C_{\vp}(t) C^{\dagger}_{\vp}(0) \ket{0^i;0^B}$.

\subsection{Similarities and differences with ref.\cite{nielsen}: }
  above and beyond confirming several results of ref.\cite{nielsen},  our study offers new and complementary results listed below that bolster the case for phenomena akin to critical dynamics for impurities with Mach number  $\beta \simeq 1$:

\vspace{1mm}

 {\textbf{i:)}The explicit dependence of the relaxation rate on the Mach number, given by  (\ref{relarateapp}),

 \vspace{1mm}

 {\textbf{ii:)} The detailed understanding of the asymptotic of dressing dynamics for $\beta >1$ confirming the modulated power law $(t_>/t)^{3/2}\,\sin(P_T t -\pi/4)$ with $t_> \propto (\beta-1)^{-2/3}$ explaining both the power and the oscillatory function with frequency $P_T \propto (\beta-1)^2$ as a consequence of a low energy \emph{threshold} in the spectral density of the impurity for $\beta > 1$.
The power $t^{-3/2}$, the       time scale  $t_> \propto (\beta-1)^{-2/3}$,  and the frequency are a direct consequence of the vanishing of the spectral density at threshold with a \emph{square root singularity}. This feature, in turn,  is a consequence of the absorption and emission of virtual long wavelength low energy phonons with linear dispersion relation.
\vspace{1mm}

 {\textbf{iii:)} The cusp   in the $\beta$-derivative,    $dz_p(\infty)/d\beta  \simeq \ln|\beta-1|$, is \emph{universal} in the sense that it is also a hallmark of the coupling to   long wavelength phonons with linear dispersion relation.

\vspace{1mm}

 \textbf{iv:)} We find that \emph{at} $\beta =1$ the asymptotic long time dressing dynamics is characterized by a power law $(t^*/t)^{1/2}$ (see Eqn. (\ref{zetapveqc})). In this case there \emph{seems} to be   a discrepancy  with the results displayed in figure (3) in ref.\cite{nielsen}.  This figure  \emph{seems} to show that the dynamics \emph{does not} reach  the steady state determined by $z_p(\infty)$ in the critical case $\beta=1$. Our study shows that such a steady state is asymptotically reached, albeit with a very slow approach $\propto t^{-1/2}$. Furthermore, we show that the $t^{-1/2}$ behavior originates in a non-analytic behavior of the spectral density at low energy at $\beta=1$ (see Eqn. (\ref{rhoveqc})). We also  find that for $\beta >1$ the asymptotic dressing dynamics, features a time scale $t_> \propto   (\beta-1)^{-2/3}$, and for $\beta <1$ a different time scale $t_< \propto (1-\beta)^{-3/2}$. However, the long time limit and the limit $\beta \rightarrow 1$ do not commute, again this is a consequence of the non-analyticity of the spectral density \emph{at} $\beta=1$ (see eqns. (\ref{kmaxgb}), (\ref{kmaxl})).

 \vspace{1mm}

    \textbf{v:)} The Weisskopf-Wigner method allows us to obtain the full quantum state   within a non-perturbative resummation of second order processes. From this state we identify the impurity Loschmidt echo and   fidelity; these are  benchmarks of quantum information. The quantum state reveals \emph{many body correlations} between the impurity and excitations of the BEC. Upon tracing over the phonon excitations we obtain the \emph{reduced} density matrix for the impurity and the von Neumann entanglement entropy, as a measure of  the correlations with and information loss into the phonon environment. We show that unitary time evolution directly relates the decay of Loschmidt echo/fidelity with the growth of entanglement entropy. We also show that there is a sharp discontinuity of the asymptotic von Neumann entanglement entropy at $\beta =1$ in weak coupling.

        The powers $t^{-3/2}$ for $\beta > 1$ and $t^{-2}$ for $\beta <1$ describing the dynamics of dressing, along with the coefficient $\propto (1-\beta)^{-3}$ in the latter case are in agreement with the results of ref.\cite{nielsen} for these cases. This agreement is, therefore, a confirmation that our approach via the spectral density is correct. Hence we conjecture that the seeming discrepancy with the results displayed in figure (3) in ref.\cite{nielsen}  in the case $\beta =1$ originates in that the  study reported in fig. (3) in this reference did not reach long enough time and seemingly did not capture the very slow dressing dynamics that our   study has shown to emerge asymptotically in this case.

        \vspace{1mm}

 \textbf{vi:)}  In the quantum master equation approach, the Born approximation entails that the impurity and phonon density matrices factorize at all times, and that the phonon density matrix does not evolve in time, namely describing the BEC vacuum state. Therefore, this approximation neglects the \emph{correlations} between the impurity and   phonon excitations. These correlations are included in the Weisskopf-Wigner approach, indeed being an integral part of the dynamical evolution. We showed how unitary time evolution relates fidelity decay to the growth of impurity-BEC correlations, therefore the emergence of these correlations is an unavoidable  consequence of unitarity.

 \vspace{1mm}

 \textbf{vii:)} The Wigner-Weisskopf state (\ref{entanstate}) is a correlated state in which the impurity is entangled with superfluid phonons. This state is remarkably similar to the variational state proposed in ref.\cite{sarma} (``Chevy'' variational state), albeit with important differences. \textbf{a:)} the coefficients in this state are completely determined by the (unitary) time evolution via the Weisskopf-Wigner equations and are not obtained from a variational principle, \textbf{b:)} unitarity constrains these coefficients to fulfill the condition $|\ma^i_{\vp}(t)|^2+ \sum_{\vk} |\ma^{iB}_{\vp,\vk}(t)|^2 =1$. For  $\beta >1$  the coefficient of the single impurity term vanishes in the long time limit.

     While seemingly both treatments yield similar results for various quantities, the Weisskopf-Wigner approach yields a direct pathway to extract the entanglement entropy and to relate its growth to the decay of the impurity fidelity. A similar feature was noticed in ref.
     \cite{boyjas} within the realm of quantum brownian motion comparing the full solution of the equations of motion with the Born approximation for the quantum master equation.

\section{Discussion.}\label{sec:discussion}

\textbf{Region of validity of approximations:} The main approximation invoked in our study above was to restrict the full dispersion relation of Bogoliubov excitations $E_k = ck\sqrt{1+(k/k^{*})^2} \rightarrow ck$. This approximation is valid for $k_{max}(\wpo) \ll k^* =2 m c$ with $k_{max}(\wpo)$ given by eqns. (\ref{kmaxgb}), (\ref{kmaxl}). As discussed in detail above, the long time dynamics is determined by the region $\wpo \simeq 0$ of the spectral density. Therefore the  most stringent constraint on the validity of the approximation arises in the case $\beta > 1$ since for $\beta \leq 1$ it follows that $k_{max}(\wpo) \rightarrow 0$ as $\wpo \rightarrow 0$. Therefore the criterion for the main approximation to be valid is given by
\be 2k_c \, (\beta-1) \ll 2 mc \Rightarrow (\beta -1) \ll r = \frac{m}{M}\,.  \label{criter}\ee Therefore the region of validity of the main approximation is larger for a heavy impurity, narrowing for an impurity that is lighter than the (bare) particles   in the BEC. Hence, for sufficiently small $|\beta-1|$ the main approximation invoked to study the long time dynamics, namely keeping solely the linear part of the Bogoliubov spectrum, is warranted.

\textbf{Wave packets:} We have studied the non-equilibrium dynamics considering that the initial state is described by a single impurity of momentum $\vp$. A more realistic scenario would generalize the single impurity state to be described by a wave packet, namely $\ket{\Psi(0)} = \sum_{\vp}\ma^i_{\vp}(0)\,\ket{1^i_{\vp};0^B}$, where now $ \ma^i_{\vp}(0)$ are the Fourier components of the single impurity wave-packet. The Weisskopf-Wigner framework can be straightforwardly adapted to this case, with the very simple modification of the initial condition, with $\ma^i_{\vp}(0)$ determined by the Fourier coefficients of the initial single particle wave packet, rather than $\ma^i_{\vp}(0)=1$ as used in this study. However, for generic wave packets the dynamical evolution becomes complicated: components with wavevectors $\vp$ obeying the Cerenkov condition will undergo relaxation along with dressing, whereas those outside the Cerenkov band will only undergo dressing dynamics. Furthermore, along with these dynamical processes, the wave packet will also undergo dispersion and spreading as in free evolution. Therefore, the full time dependence of a wave packet will exhibit complicated dynamics that will be the result of all the different processes.

\textbf{Cascade decay:} We have considered a weak coupling between the impurity and the BEC, and  the vertex describing the absorption and/or emission of a superfluid phonon is associated with one power of the effective dimensionless coupling $\lambda M$. In weak coupling, the leading order self-energy thus describes the process in which the emission of a phonon is followed by its absorption (see fig. (\ref{fig:selfenergy})), hence the leading order self-energy is of order $\lambda^2 M^2$.  To this leading order the impurity with $\vp$ emits a phonon with $\vk$, thus leaving the impurity with $\vp-\vk$. However if $|\vp-\vk|> Mc$, the momentum of the impurity can still satisfy the Cerenkov condition and continue to emit ``on-shell'' phonons in a \emph{cascade}: $\vp \rightarrow \vp-\vk_1 \rightarrow \vp-\vk_1-\vk_2 \rightarrow \cdots $. Each step in this cascade results in one further vertex and one more power of $\lambda M$. The general form of the time evolved quantum state during the cascade is
\be \ket{\Psi(t)}  = \ma^i_{\vp}(t)\,\ket{1^i_{\vp};0^B} + \sum_{\vk} \ma^{iB}_{\vp,\vk}(t) \ket{1^i_{\vp-\vk};1^B_{\vk}}+ \sum_{\vk;\vq} \ma^{iB}_{\vp,\vk,\vq}(t) \ket{1^i_{\vp-\vk-\vq};1^B_{\vk};1^B_{\vq}} + \cdots  \,, \label{cascade} \ee where the amplitudes are obtained from the hierarchy of Weisskopf-Wigner equations: $\ma^{iB}_{\vp,\vk}(t)\propto \lambda M\,;\, \ma^{iB}_{\vp,\vk,\vq}(t) \propto (\lambda M)^2 \,;\,\mathrm{etc}$.

For strong impurity-BEC coupling the impurity will emit (and absorb) phonons in a cascade diminishing its momentum via Cerenkov phonon emission.
 The consistent analysis of the dynamics in this case requires a systematic \emph{resummation} of the events described by this cascade, and  the self-energy will include rainbow and crossed vertex corrections so that the resulting excited state will be a superposition of multi-phonon states.  A systematic study of this case is a worthy endeavor; however, it is well beyond the scope of this article.

\textbf{$\beta_f \leq 1$ as a dynamical attractor manifold?} In this article we focused on studying the non-equilibrium dynamics for $\beta \simeq 1$, revealing a slow-down of relaxation  and dressing in agreement with ref.\cite{nielsen}. It would seem that restricting the study to this region of Mach number implies a \emph{fine tuning} of initial conditions. However the combination of the wave packet and cascade arguments above, when combined with the results obtained in our study suggest that an impurity quenched into a BEC with a   initial Mach number $\beta_i \gg 1$   will relax, asymptotically, to a Mach number $\beta_f \leq 1$ by multiphonon Cerenkov emission. To understand the main arguments behind this \emph{conjecture}, let us consider the quantum state (\ref{entanstate}) obtained in leading  order, namely the first two contributions to the quantum state (\ref{cascade}). The one-phonon state in (\ref{entanstate}) (or in (\ref{cascade})) can be interpreted as a \emph{wave packet}, since it is a linear superposition of states with wavevectors $\vk$ with Fourier coefficients given by $\ma^{iB}_{\vp,\vk}(t)$.  As discussed above, the amplitude $\ma^{iB}_{\vp,\vk}(t)$ is largest for the band of wavevectors $\vk$ that satisfy the Cerenkov condition, and  among these,  it is largest for $k_{max}= 2k_c(\beta-1)$ because the matrix element is largest for largest value of $k$. If the initial value of the Mach number of the impurity is $\beta_i = 1+ \delta$ with $\delta \ll 1$, it follows that the typical value of $\beta_f$ for the impurity in the one-phonon state in  (\ref{entanstate}) is $\beta_f =|\vp-\vk|/Mc \simeq 1-\delta$ (with $k\simeq k_{max}$). Therefore, this one-phonon state contribution to the full quantum state is \emph{stable} under the higher order process of Cerenkov emission with a two- phonon final state. However, if $\beta_i \gg 1$ there will be components with wavevector $\vk$ in the one-phonon \emph{wave packet} such that $\beta_f > 1$. Including the higher order transition in the Weisskopf-Wigner hierarchy, to the two-phonon state, will allow these components to \emph{decay} via Cerenkov emission of another phonon on a longer time scale, since the transition probability is suppressed by   two more powers of the coupling. Therefore, asymptotically at long time the coefficients $\ma^{iB}_{\vp,\vk}(\infty)$ will be non-vanishing \emph{only} for those wavevectors $\vk$ for which $\beta_f \leq 1$, namely $|\vp-\vk|/Mc \leq 1$. The relaxation of this coefficient to its asymptotic value will be on much longer time scales $\propto  (\lambda M)^{-4}\times (\beta_f-1)^{-\kappa}$ with $\kappa$ a positive integer that depends on the multi-phonon phase space. Generalizing this argument to multi-phonon states, we are led to \emph{conjecture} that the \emph{exact} form of the asymptotic quantum state for an impurity that was initially quenched with $\beta_i \gg 1$ is given by
\be \ket{\Psi(\infty)}  =   \sum_{\vk} \ma^{iB}_{\vp,\vk}(\infty) \ket{1^i_{\vp-\vk};1^B_{\vk}}+ \sum_{\vk;\vq} \ma^{iB}_{\vp,\vk,\vq}(\infty) \ket{1^i_{\vp-\vk-\vq};1^B_{\vk};1^B_{\vq}} + \cdots \,,\label{asycascade}\ee where the amplitudes $ \ma^{iB}_{\vp,\vk}(\infty)\,;\,\ma^{iB}_{\vp,\vk,\vq}(\infty)\,;\,\mathrm{etc}$ are non-vanishing \emph{only} for those wave vectors for which the Mach number of the impurity in the corresponding state is $\beta_f = |\vp-\vk-\vq -\cdots|/Mc \leq 1$. Unitarity implies that $ \sum_{\vk} |\ma^{iB}_{\vp,\vk}(\infty)|^2 + \sum_{\vk;\vq} |\ma^{iB}_{\vp,\vk,\vq}(\infty)|^2   + \cdots  = 1$. These arguments lead us to the main \emph{conjecture}:  that for asymptotically long time, the full quantum \emph{steady state} will be a linear superposition of the impurity with entangled multi-phonon states of the form given by Eqn. (\ref{asycascade})  in which the Mach number of the impurity is always $\leq 1$. This asymptotic state is, therefore, a dynamical attractor. The results obtained  for $\beta \simeq 1$ indicate  that this steady-state attractor will be reached at asymptotically long time because as the effective Mach number becomes smaller the relaxation via the emission of the next Cerenkov phonon occurs on a longer time scale. This analysis provides a quantum many body interpretation of the ``classical intuitive argument'', based on the Landau criterion for superfluidity, that a supersonic mobile impurity will slow down by creating excitations until it  becomes subsonic. Eqn. (\ref{asycascade}) is the many body quantum state that describes this asymptotic state.

\textbf{Trapped cold atoms:} We have studied the non-equilibrium dynamics of polaron relaxation and dressing in the case of a homogeneous BEC. Experimentally most of the studies of many body physics of BEC are performed with harmonically trapped atoms, therefore the spectrum of low energy excitations above the condensate is very different from the homogeneous case. In particular the process of Cerenkov radiation of phonons, which is the main relaxational channel for a mobile impurity in the homogeneous case will be substantially modified, if relevant at all, in the case of trapped cold atoms. Therefore, while the results of our study may be a prelude towards understanding the non-equilibrium many body dynamics of impurities in a condensate, their applicability to the typical experimental setup with trapped cold atoms would have to be reassessed. Recently the non-equilibrium   dynamics of a \emph{subsonic}  impurity immersed in a \emph{one dimensional trapped BEC} has been reported\cite{oned}. A numerical solution of the Schroedinger equation in this reference reveals intriguing dynamics of transfer of energy from the impurity to the bath. Although the setting and the initial condition differ from our study, the energy transfer between impurity and bath degrees of freedom seems to be physically similar to the relaxation dynamics that we studied. It is definitely a worthy endeavor to extrapolate the methods and results obtained in our study to the case of harmonically trapped gases.

\section{Conclusion and further questions. }\label{sec:conclusions}

Motivated by the fundamental importance of the polaron as a paradigm of a quasiparticle in many body physics, and by current  experiments that  access its  relaxation and dressing dynamics with unprecedented control, we   introduced an alternative method to study its non-equilibrium time evolution. We consider an impurity suddenly immersed --quenched-- with velocity $v$ into the ground state of an homogeneous Bose Einstein condensate at zero temperature. A many body generalization of the Weisskopf-Wigner method allows us to study the dynamics of the Loschmidt echo or fidelity (survival probability) of the impurity  along with the emergence of  \emph{correlations} between the impurity and low energy excitations of the (BEC).  A wealth of dynamical scales characterize the dynamics of relaxation and polaron formation.   Early time transient dynamics feature quantum Zeno behavior crossing over to a stretched exponential. The Mach number of the impurity $\beta = v/c$ with $c$ the speed of sound of superfluid phonons plays  a crucial role in the characterization of the intermediate and asymptotically long time dynamics. The region $\beta \simeq 1$ features  a slowing down of relaxation and dressing dynamics in agreement with results  previously reported in ref.\cite{nielsen}.  For $\beta > 1$ the fidelity \emph{decays} via Cerenkov emission of long wavelength phonons with a rate $\Gamma_p \propto (\beta-1)^3$, whereas dressing dynamics is characterized by a power law $\propto t^{-3/2}$ on a time scale $t_> \propto (\beta-1)^{-2/3}$ modulated by oscillations with frequecy $\propto (\beta-1)^2$ as a consequence of emission and absorption of ``off-shell'' phonons with virtuality $\propto (\beta-1)^2$. For $\beta \leq 1$ only the process of dressing is available, resulting in power law behaviors for the polaron residue. For $\beta =1$ we find that the residue approaches its asymptotic value with a power law $t^{-1/2}$, whereas for $\beta < 1$ as $(t_</t)^{-2}$  on a time scale   $t_<\propto (1-\beta)^{-3/2}$. The sharp change in asymptotic behavior as $\beta$ crosses $\beta =1$ is a consequence of the non-analyticity of the impurity spectral density for $\beta \simeq 1$. These results show that relaxation and dressing dynamics undergo slowing down as $\beta \rightarrow 1$. The asymptotic value of the polaron residue features a cusp $\simeq \ln|1-\beta|$  in its $\beta$-derivative at $\beta =1$. These features, namely  the $\beta$ dependence of $\Gamma_p$, the power law exponents along with their pre-factors,   the oscillatory modulation of the dressing dynamics  and the cusp in the $\beta$- derivative of the polaron residue at $\beta =1$ are all a hallmark of absorption and emission of long wavelength superfluid phonons with linear dispersion relation.

We obtain the impurity reduced density matrix from the time evolution of the initial state, and its entanglement entropy, as a measure of information loss and of \emph{correlations} between the impurity and excitations in the BEC. We show how unitarity directly relates the fidelity decay of the impurity, either via relaxation or dressing, to the growth of the entanglement entropy. We find a sharp transition in the asymptotic entanglement entropy at $\beta =1$.

Taken together, the slowing down of relaxation and dressing featuring power laws in time,  with time scales that diverge as $\beta \rightarrow 1$, along with the sharp discontinuity in   the entanglement entropy   and the $\beta$ derivative of the polaron residue,     suggest  \emph{universal} dynamical  critical  phenomena featuring a  slowing down of non-equilibrium dynamics for $\beta \simeq 1$. These phenomena are a direct consequence of emission and absorption of long-wavelength superfluid phonons with linear dispersion relation.

We have \emph{conjectured} that for an impurity quenched into a BEC with $v\gg c$, the non-equilibrium time evolution leads to an asymptotic steady state dynamical attractor with effective Mach number $\beta_f \leq 1$ via a cascade process resulting in the impurity being entangled with multi-phonon states. This attractor will be reached at asymptotically long time scales as a consequence of the slowing down of relaxation as the effective $\beta \rightarrow 1$.

We have also established a relation between the quantum state obtained from unitary time evolution via the Weisskopf-Wigner method and a variational state for a Bose polaron proposed in ref.\cite{sarma} inspired by a similar variational state for a Fermi polaron introduced in ref.\cite{chevy}.

   Further understanding of the nature of dynamical critical phenomena for $\beta \simeq 1$, along with a systematic method to analyze the cascade processes  yielding multi-phonon states and the asymptotic attractor, merits  deeper study implementing non-perturbative methods.

\appendix
\section{Unitarity.}\label{app:unitarity}

Unitarity is equivalent to conservation of probability. The set of   equations (\ref{eofm}) for the coefficients $\ma_n(t)$, along with the hermiticity  of $H_I(t)$  lead to
\be \frac{d}{dt} \sum_{n}\,|\ma_n(t)|^2 = -i \sum_{m,n} \Big[\ma_m(t)\,\ma^*_n(t)\,\bra{n}H_I(t)\ket{m} - \ma_n(t)\,\ma^*_m(t)\,\bra{m}H_I(t)\ket{m} \Big] \,.  \label{derimod}\ee Relabelling $m \leftrightarrow n$ one finds
\be  \frac{d}{dt} \sum_{n}\,|\ma_n(t)|^2 = 0\,, \label{constprob}\ee namely the sum of probabilities is time independent. Setting  $\ma^i_{\vp}(t=0) =1$ and all other coefficients to vanish at the initial time, yields Eqn. (\ref{unitarity1}).

\section{Spectral density:}\label{app:specdens}

Changing integration variable in the angular integral in (\ref{rhoexp}) to $ z= vk\cos(\theta)$ yields
\be \rho(\wpo;p)= \frac{\lambda^2}{v}\,\int^\infty_0 \frac{k^2}{\sqrt{1+\frac{k^2}{k^{*2} }}}\,\int^{vk}_{-vk}dz \,\delta\Big[z-\Big(\frac{k^2}{2M}+E_k-
\widetilde{p}_0\Big) \Big]\, dk ~~;~~ \wpo = p_0 -\epsilon_p \,. \label{newrho}\ee The z-integral yields
\be \int^{vk}_{-vk}dz \,\delta\Big[z-\Big(\frac{k^2}{2M}+E_k-
\widetilde{p}_0\Big) \Big] = \Bigg\{\begin{array}{l}
                                      1~~\mathrm{if} ~~ -vk \leq \frac{k^2}{2M}+ck\sqrt{1+\frac{k^2}{{k^*}^2}}-
\widetilde{p}_0 \leq vk \\
                                      0 ~~\mathrm{otherwise}
                                    \end{array}
 \label{zint} \ee
  The momentum region within which the delta function in (\ref{zint}) is satisfied  results in the constraint $k_{min} \leq k \leq k_{max}$ where $k_{min},k_{max}$ are the intersections of the curve $\frac{k^2}{2M}+ck\sqrt{1+\frac{k^2}{{k^*}^2}}-
\widetilde{p}_0$ with the straight lines $\pm vk$ respectively.

 These are depicted in figures (\ref{fig:intersectionmin},\ref{fig:intersectionplus}) for $\wpo >0$ and $\wpo<0$ respectively.

\begin{figure}[ht!]
\begin{center}
\includegraphics[height=4in,width=4in,keepaspectratio=true]{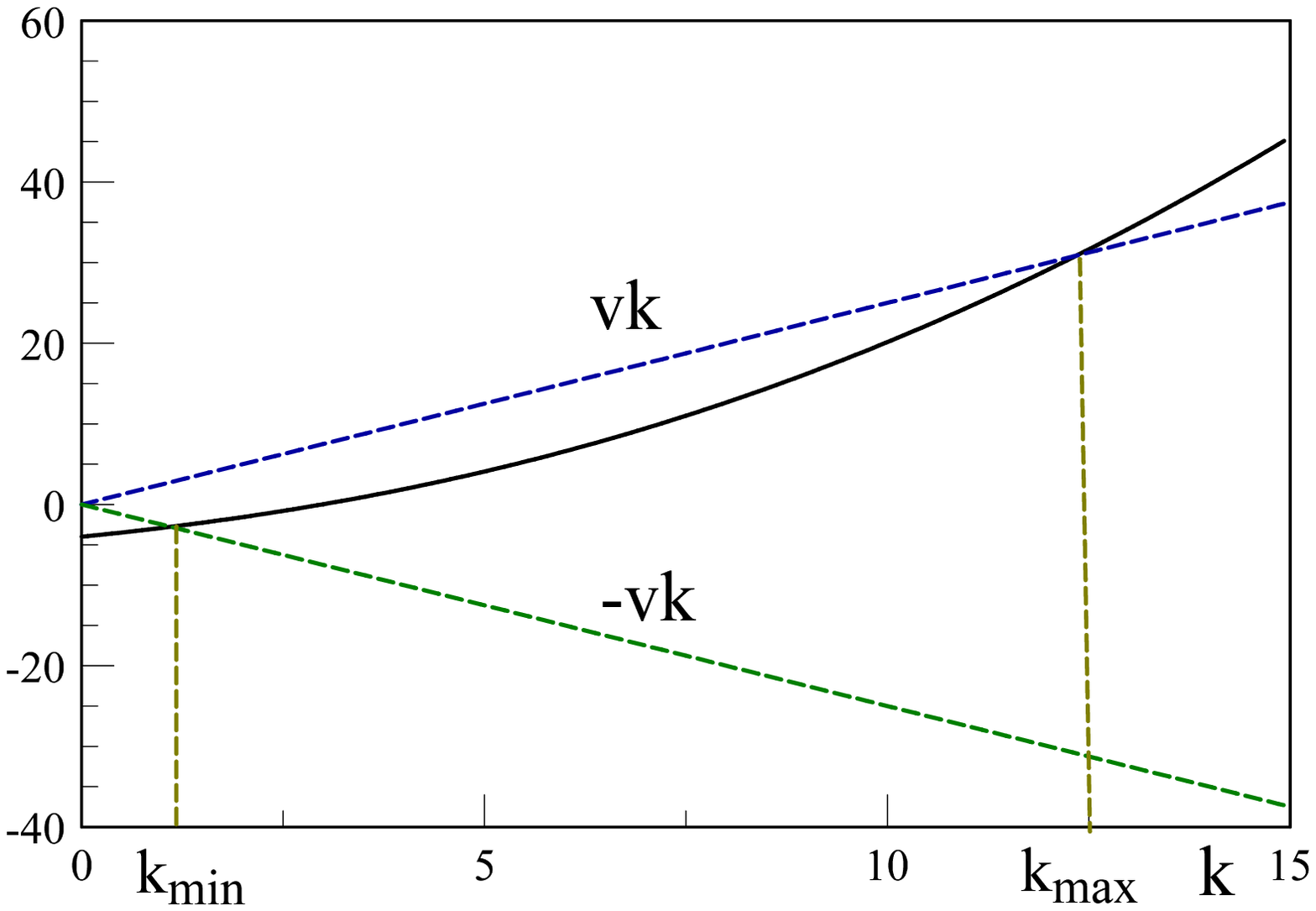}
\caption{Solutions for the constraints Eqn. (\ref{zint}) for $\wpo >0$. The solid line is the
curve $\frac{k^2}{2M}+ck\sqrt{1+\frac{k^2}{{k^*}^2}}-
\widetilde{p}_0$ . }
\label{fig:intersectionmin}
\end{center}
\end{figure}

\begin{figure}[ht!]
\begin{center}
\includegraphics[height=4in,width=4in,keepaspectratio=true]{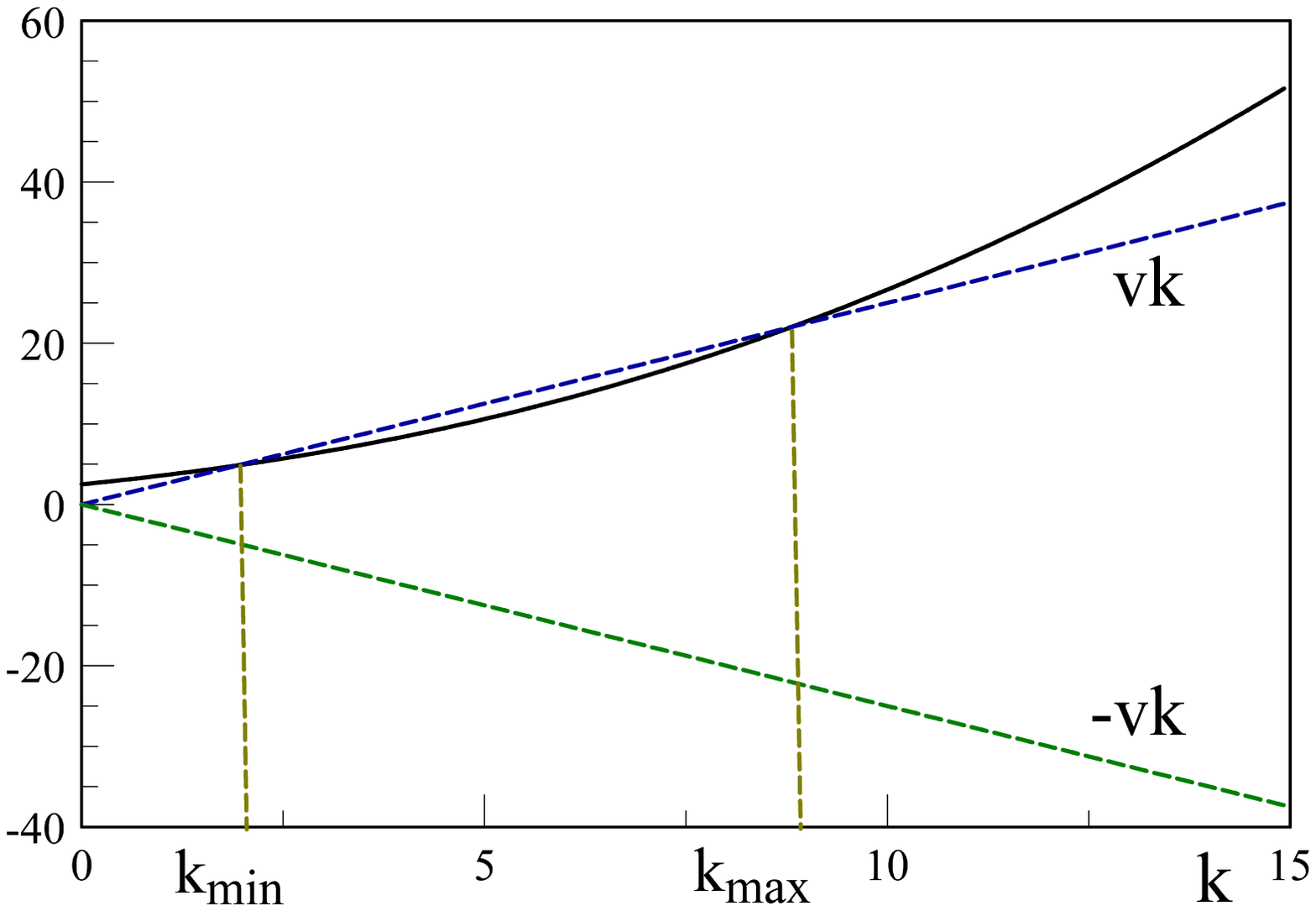}
\caption{Solutions for the constraints Eqn. (\ref{zint})  for $\wpo<0$.  The solid line is the
curve $\frac{k^2}{2M}+ck\sqrt{1+\frac{k^2}{{k^*}^2}}-
\widetilde{p}_0$ }
\label{fig:intersectionplus}
\end{center}
\end{figure}

Once the values of $k_{min},k_{max}$ are established, the k-integral in the region $k_{min}\leq k \leq k_{max}$ is straightforward, and we find
\be \rho(\wpo,p) = \frac{\lambda^2}{2v}\,(k^*)^3 \Bigg\{\frac{k_{max}}{k^*}\,F[k_{max}]-\frac{k_{min}}{k^*}\,F[k_{min}] - \ln\Bigg[ \frac{\frac{k_{max}}{k^*}+F[k_{max}]}{\frac{k_{min}}{k^*}+F[k_{min}]} \Bigg]  \Bigg\} ~~;~~ F[k] = \sqrt{1+\frac{k^2}{{k^*}^2}  } \label{rhofina}\ee

For $\wpo < 0$ the curve \emph{only} intersects the straight line $+vk$  and \emph{only} for $v > c$, there is a \emph{minimum} value of $\wpo$ below which there is no further solution. This minimum value defines a \emph{threshold}, at which $k_{min}=k_{max}$. This behavior is displayed in fig. (\ref{fig:intersectionplus}).  For $v<c$ there is no longer intersection with the $+ vk$ line, hence there is no solution for $\wpo <0; v< c$.

 For $\wpo >0$ the curve  intersects both   straight lines $\pm vk$ as shown in fig. (\ref{fig:intersectionmin}).

 \vspace{1mm}

 \textbf{Large $\wpo$ limit:}

 The large $\wpo$ limit of the spectral density can be extracted in a simple manner: in the original form, Eqn.  (\ref{rhoexp}), we can neglect $v$, furthermore, for large $\wpo$  the delta function will be satisfied for large $k$, therefore take $k \gg k^*$, hence $E_k = k^2/2m$. Now the integrals are straightforward and yield the asymptotic limit
 \be \rho(\wpo,p) = \lambda^2 k^* (2\mu)^{3/2} \,\sqrt{\wpo} ~~;~~ \mu = \frac{mM}{m+M} \label{largeporho}\ee  for $\mu \, \wpo \gg (k^*)^2 $.

 Finding the roots $k_{max},k_{min}$ for arbitrary values of $\wpo$ and parameters in general must be done numerically, however, as discussed in the main text, the long time limit is determined by the small $\wpo$ region of $\rho$. Furthermore, we also focus on the ``critical'' region $v \simeq c$, as discussed above, this region is dominated by the emission and absorption of long wavelength phonons whose dispersion relation is   $E_k = ck$. The equations that determine the intersections $k_{max},k_{min}$   in this case simplify to,
 \be \frac{k^2}{2M}+ck-\wpo = \pm v\,k \,, \label{intersmall} \ee and the acceptable  solutions must correspond to $k>0$.

 It is convenient to introduce the following energy  and momentum scales and Mach number $\beta$,
 \be \epsilon_c = \frac{1}{2} M\,c^2 ~~;~~ k_c = M\,c ~~;~~ \beta = \frac{v}{c}\,, \label{params}\ee in terms of which we find  the following solutions

 \vspace{2mm}

 \textbf{A:) $\beta > 1$  }

 \textbf{i:)} $\wpo >0$: Intersection with both branches $\pm v\,k$,

 \bea k_{max} & = &  k_c \Bigg[ (\beta-1) + \sqrt{(\beta-1)^2 + \frac{\wpo}{\epsilon_c}} ~\Bigg] \label{kmaxgapp} \\
 k_{min} &  = &  k_c \Bigg[ -(\beta+1) + \sqrt{(\beta+1)^2 + \frac{\wpo}{\epsilon_c}} ~ \Bigg] \label{kmingapp} \eea

 \textbf{ii:)} $\wpo <0$: Intersection only with branch $+v\,k$

 \bea  k_{max} & = &  k_c \Bigg[ (\beta-1) + \sqrt{(\beta-1)^2 + \frac{\wpo}{\epsilon_c}} ~\Bigg] \label{kmaxlapp} \\
  k_{min} & = &  k_c \Bigg[ (\beta-1) - \sqrt{(\beta-1)^2 + \frac{\wpo}{\epsilon_c}} ~\Bigg] \label{kminlapp} \,. \eea
  For $v>c, \wpo <0$ the solutions (\ref{kmaxl},\ref{kminl}) are only available for $-\epsilon_c \, (\beta-1)^2 \leq \wpo$. This is evident from the expressions (\ref{kmaxlapp},\ref{kminlapp}) in which the argument of the square roots are positive only for $-\epsilon_c\,(\beta-1)^2 \leq \wpo$. This \emph{minimum value} of $\wpo$ determines a lower \emph{threshold} for the spectral density at which $k_{max}=k_{min}$ at which point $\rho(\wpo;p)=0$, therefore the region of \emph{negative} $\wpo$ for which the spectral density has support is given by

  \be -P_T \leq \wpo ~~;~~ P_T = \epsilon_c \,(\beta-1)^2 \,.  \label{threshold} \ee

  This threshold can also be understood from fig. (\ref{fig:intersectionplus}): the dotted straight line $+vk$ becomes the \emph{tangent} to the solid line (the curve $\frac{k^2}{2M}+ck\sqrt{1+\frac{k^2}{{k^*}^2}}-
\widetilde{p}_0$)  when $k_{max} = k_{min}$, which from the expressions (\ref{kmaxlapp},\ref{kminlapp}) (in the long wavelength limit) corresponds to $\wpo = -P_T$. For $\wpo < -P_T$ there are no further solutions and the spectral density vanishes.

 \vspace{2mm}

\textbf{B:) $\beta < 1$  }

There are no intersections for $\wpo <0$, there are solutions \emph{only} for $\wpo >0$, these are given by ($\wpo>0$ only)
  \bea k_{max} & = &  k_c \Bigg[ (\beta-1) + \sqrt{(\beta-1)^2 + \frac{\wpo}{\epsilon_c}} ~\Bigg] \label{kmaxgbapp} \\
 k_{min} &  = &  k_c \Bigg[ -(\beta+1) + \sqrt{(\beta+1)^2 + \frac{\wpo}{\epsilon_c}} ~ \Bigg] \,. \label{kmingbapp} \eea

 For the ``near critical'' case $\beta \simeq 1$ and small $\wpo$, which is the relevant region for the long time dynamics, the spectral density simplifies substantially, we find in this case

 \be \rho(\wpo;p) = \frac{\lambda^2}{3 \beta \,c }\,\Bigg[\Big(k_{max}(\wpo)\Big)^3 -  \Big(k_{min}(\wpo)\Big)^3 \Bigg] = \frac{2\lambda^2 M^2\,\epsilon_c}{3\beta} \,\Bigg[\Big(\frac{k_{max}(\wpo)}{k_c}\Big)^3 -  \Big(\frac{k_{min}(\wpo)}{k_c}\Big)^3 \Bigg]\,. \label{rhosmallsapp}\ee

 \section{$z_p(\infty)$:}\label{app:zinfinity}
The asymptotic value $z_p(\infty)$ is given by Eqn. (\ref{longtigam}), with the definition of the spectral density, Eqn. (\ref{rhoexp}), the integral over $\wpo$ yields,
\be \mathcal{P} \int^\infty_{-\infty} d\wpo \, \frac{ \rho(\wpo,p)}{\wpo^{\,2}}  = \lambda^2 \mathcal{P}  \int^\infty_0 \,dk\, \frac{k^3}{\sqrt{1+\frac{k^2}{k^{*2}}}}\,\int^1_{-1}d(\cos(\theta))\,\frac{1}{\Big(\frac{k^2}{2M}+E_k-
v\,k\,cos(\theta)\Big)^2}\,. \label{intC1} \ee Writing
\be  \frac{1}{\Big(\frac{k^2}{2M}+E_k-
v\,k\,cos(\theta)\Big)^2} = \frac{1}{vk}\, \frac{d}{d(\cos(\theta))}
\,\Bigg[\frac{1}{\Big(\frac{k^2}{2M}+E_k-
v\,k\,cos(\theta)\Big)}\Bigg]\,   \ee the angular integral is straightforward. Introducing the following variables
\be x \equiv \frac{k}{k^*} ~~,~~  r = \frac{m}{M} \ee we find
\be z_p(\infty)   = \frac{\lambda^2 {k^*}^2}{\beta c^2}\,\mathrm{Re}\, \Bigg\{\int^\infty_0 dx\, \frac{x}{\sqrt{1+x^2}}\,\,  \frac{1}{rx+\sqrt{1+x^2}-\beta -i\eta} - (\beta \rightarrow -\beta)  \Bigg\}~~;~~ \eta \rightarrow 0^+ \label{zfinint}\,. \ee A closed form can be obtained for the equal mass case $r=1$ for which $k^*=2 Mc$: changing variables to $x = \frac{1}{2}(y-\frac{1}{y})$,   we   find in this case
\be z_p(\infty) = \Big[\frac{2\lambda M}{\beta } \Big]^2\,\,\Bigg[  \frac{\beta^2-1}{2\beta} ~\ln \Bigg| \frac{1+\beta}{1-\beta}\Bigg| +1\Bigg] \,. \label{zfinir1app} \ee

\section{Long time behavior for $v>c$:}\label{app:ltvlarc}

As explained in the text, we first calculate the time dependent $\emph{rate}$, $\dot{\gamma}_p(t)$ and integrate back in time with the boundary condition that $\gamma_p(t) \rightarrow \Gamma_p t + z_p(\infty)$ as $t \rightarrow \infty$. Introducing explicitly the lower threshold $P_T=\epsilon_c\,(\beta-1)^2$, it follows that the rate is given by
\be \dot{\gamma}_p(t) = \int^\infty_{-P_T} \rho(\wpo;p) \, \frac{\sin(\wpo t)}{\wpo}\, d\wpo\,.   \label{dotgama} \ee We now write $\rho(\wpo;p) \equiv \rho(0;p)+ \overline{\rho}^{>}(\wpo;p)+\overline{\rho}^{<}(\wpo;p)$, where $\overline{\rho}^{\gtrless}$  only feature support for $\wpo \gtrless0 $ respectively, yielding

\be \dot{\gamma}_p(t) = \underbrace{\rho(0;p)\,\int^\infty_{-P_T t} \frac{\sin(x)}{x} dx}_{(a)} + \underbrace{\int^\infty_0 \overline{\rho}^{>}(\wpo;p) \,\frac{\sin(\wpo t)}{\wpo} \,d\wpo}_{(b)} +
\underbrace{\int^0_{-P_T} \overline{\rho}^{<}(\wpo;p) \,\frac{\sin(\wpo t)}{\wpo} \,d\wpo }_{(c)}\, \label{gamasplit2}\ee

\vspace{1mm}

\textbf{(a):}

\be \int^\infty_{-P_T t} \frac{\sin(x)}{x} dx = 2 \int^\infty_{0} \frac{\sin(x)}{x}\, dx - \int^\infty_{P_T t} \frac{\sin(x)}{x} dx = \pi - \frac{\cos(P_T t)}{P_T t} + \cdots \label{sin1}\ee therefore
\be (a) = \pi \rho(0;p) \Bigg[ 1- \frac{\cos(P_T t)}{\pi P_T t} + \cdots \Bigg]\,.  \label{terma} \ee

\vspace{1mm}

\textbf{(b):}  write
\be \sin(\wpo t) = -\frac{1}{t} \frac{d}{d\wpo} \cos(\wpo t) ~~;~~  \cos(\wpo t) = \frac{1}{t} \frac{d}{d\wpo} \sin(\wpo t)\, \label{appequ}  \ee and integrate by parts.

For $\beta >1$ the spectral density is analytic in $\wpo$ as $\wpo \rightarrow 0$, it is given by
\be \rho(\wpo;p) = \rho(0;p) + \rho'(0;p)\,\wpo + \cdots \label{smallpo} \ee with $\rho'(0;p) = d\rho(\wpo;p)/d\wpo\,|_{\wpo=0}$ and the dots stand for higher integer powers of $\wpo$. After  integration by parts  using (\ref{appequ}),   using the large and small $\wpo$   behavior (\ref{largeporho},\ref{smallpo}), respectively  and subtracting $\rho(0;p)$, we find

 \be (b) = \frac{1}{t} \, \rho'(0;p) + \mathcal{O}(1/t^2)\,.  \label{bterm}\ee

 \vspace{1mm}

 \textbf{(c):} Use   (\ref{appequ}) to find
 \be (c)= -\frac{1}{t} \Bigg[\frac{\overline{\rho}^{<}(\wpo;p)}{\wpo}\,\cos(\wpo t) \Bigg]^{0}_{-P_T}+ \frac{1}{t}\,  \int^0_{-P_T} \frac{d}{d\wpo}\Bigg(\frac{\overline{\rho}^{<}(\wpo;p)}{\wpo}\Bigg) \,\cos(\wpo t) \, d\wpo\,, \label{cterm1} \ee the first term yields
 \be -\frac{1}{t} \, \rho'(0;p) + \frac{\rho(0;p)}{P_T t} \cos(P_T t) \label{c1}\ee where we used that $\rho(-P_T;p)=0$ because for $\wpo <0$ it follows from (\ref{kmaxlapp}, \ref{kminlapp} ) that $k_{max}(-P_T) = k_{min}(-P_T)$, therefore $\overline{\rho}(-P_T;p) = -\rho(0;p)$. Combining the above results for the contributions (a,b,c), we find

 \be \dot{\gamma}_p(t) = \pi \rho(0;p) + \underbrace{\frac{1}{t} \, \int^0_{-P_T} \frac{d}{d\wpo}\Bigg(\frac{\overline{\rho}^{<}(\wpo;p)}{\wpo}\Bigg) \,\cos(\wpo t) \, d\wpo}_{(D)}\,. \label{dotgamfins}
  \ee

  In the region near the upper limit $\wpo \simeq 0$, it is straightforward to find that the spectral density features an analytic power series expansion $\overline{\rho}(\wpo;p) = A \wpo + B\wpo^2 +\cdots$, hence upon rescaling $\wpo t \rightarrow x$ this region yields inverse integer powers of $t$: $1/t^2 ; 1/t^3 \cdots$. In the region near the lower limit $\wpo \simeq -P_T$ it follows from eqns. (\ref{kmaxlapp},\ref{kminlapp}) and (\ref{rhosmallsapp}), that
  \be \rho^<(\wpo;p) \simeq \frac{\lambda^2\,k^3_c}{3 \beta \,c }   \Bigg[ 6\,(\beta-1)^2 \,\Big({\frac{P_T}{\epsilon_c}+\frac{\wpo}{\epsilon_c}}\Big)^{1/2} + 2\,\Big({\frac{P_T}{\epsilon_c}+\frac{\wpo}{\epsilon_c}}\Big)^{3/2}  \Bigg] \,,\label{rholesspt} \ee therefore near threshold we find
  \be \frac{d}{d\wpo}\Bigg(\frac{\overline{\rho}^{<}(\wpo;p)}{\wpo}\Bigg) = \frac{\lambda^2\,k^3_c}{ \beta \,c }\, \frac{(\beta-1)^2}{\wpo\,\sqrt{\epsilon_c}}\,\frac{1}{\sqrt{P_T+\wpo}} \Big[ 1 + \cdots \Big] \label{dernearpt}\ee where the dots stand for \emph{positive integer} powers of $(P_T+\wpo)$. The first term in (\ref{rholesspt}) yields the leading contribution from the region near threshold in the long time limit. The square root singularity at $\wpo \simeq -P_T$ yields a fractional power of time in the long time limit, which is the leading asymptotic behavior. To see this insert (\ref{dernearpt}) into the integral in (\ref{dotgamfins}) and rescale $(P_T+\wpo)t \equiv x$, the leading contribution to the  integral (D) in the long-time limit becomes
  \be (D) = - \frac{1}{t^{3/2}}\, \frac{\lambda^2\,k^3_c}{ \beta \,c }\, \frac{(\beta-1)^2}{P_T\, \sqrt{\epsilon_c}} \Bigg\{ \cos(P_T t) \,\int^{P_T t}_0 \frac{\cos(x)}{(1-\frac{x}{P_T\,t})\sqrt{x}}\,dx +\sin(P_T t) \,\int^{P_T t}_0 \frac{\sin (x)}{(1-\frac{x}{P_T\,t})\sqrt{x}}\,dx \Bigg\} \,,\label{leadorthre}\ee  with corrections which are higher inverse  powers of $t$. Now taking the $P_T \, t\rightarrow \infty$ limit in the integrals and carrying out the remaining integrals in $x$ we finally find
  \be (D) = -\frac{\sqrt{\pi}\,\lambda^2\,k^3_c}{ \beta \,c } \, \frac{\cos[P_T t - \pi/4]}{(\epsilon_c\, t)^{3/2} } \,. \label{finathre}\ee Gathering the results obtained above up to leading order we find
  \be \dot{\gamma}_p(t) = \frac{\Gamma_p}{2}-\frac{\sqrt{\pi}\,\lambda^2\,k^3_c}{ \beta \,c } \, \frac{\cos[P_T t - \pi/4]}{(\epsilon_c\, t)^{3/2} } + \mathcal{O}(1/t^2)\,.  \label{dotgamafinish}\ee This result is valid for $P_T t \gg 1$.

  The next step is to integrate this result in time and append the asymptotic boundary condition. The integral in time of the oscillatory term cannot be found in a useful closed form, it being related to a Fresnel integral. However, progress can be made in the asymptotic long time limit with $P_T t \gg 1$ by writing
\be \cos[P_T t - \pi/4] = \frac{1}{P_T} \,\frac{d}{dt} \sin[P_T t - \pi/4]\,, \label{ide7}\ee and integrating  by parts, a process that can be iterated to yield
\be \int^t \frac{\cos[P_T t' - \pi/4]}{ {t'}^{\,3/2} } dt' =  \frac{1}{P_T t^{3/2}}\Bigg[
\sin[P_T t- \pi/4] - \frac{3}{2P_T t}\,\cos[P_T t- \pi/4] +\cdots\Bigg] \,,  \label{asyosciint}\ee where the dots stand for higher  powers of $1/(P_T t)$ and the integration constant is fixed by the boundary condition in the asymptotic long time limit. Therefore, for $P_T t \gg 1$ we obtain
\be \gamma_p(t) = \frac{\Gamma_p}{2}\, t + z_p(\infty) - \Bigg[\frac{t_>}{t} \Bigg]^{3/2}\,\sin[P_T t- \pi/4] + \mathcal{O}(1/t^2)+\cdots ~~;~~ t_> = \frac{1}{\epsilon_c}\,\Bigg[\frac{ \sqrt{4\pi}\, \lambda^2 M^2}{\beta\,(\beta-1)}\Bigg]^{2/3}\,. \label{gamathreshold}\ee

\end{document}